\documentclass[11pt,a4paper]{article}
\usepackage[T1]{fontenc}
\usepackage[utf8]{inputenc}
\usepackage{lmodern}
\usepackage[margin=2.7cm]{geometry}
\usepackage{amsmath,amssymb,amsthm,mathtools}
\usepackage{booktabs,tabularx}
\usepackage{tikz}
\usepackage[hidelinks]{hyperref}
\hypersetup{pdftitle={Cyclotomic Resonances and Generic Non-D-finiteness of the Three-Dimensional Ising Model},pdfauthor={Z. C. Tu}}
\usepackage{microtype}
\numberwithin{equation}{section}
\newtheorem{theorem}{Theorem}[section]
\newtheorem{lemma}[theorem]{Lemma}
\newtheorem{proposition}[theorem]{Proposition}
\newtheorem{corollary}[theorem]{Corollary}
\theoremstyle{definition}

\newtheorem{remark}[theorem]{Remark}
\newcommand{\Q}{\mathbb Q}
\newcommand{\C}{\mathbb C}
\newcommand{\Z}{\mathbb Z}
\newcommand{\Id}{\mathbf 1}
\newcommand{\Nop}{\widehat N}
\newcommand{\dd}{\mathrm d}
\newcommand{\ii}{\mathrm i}

\newcommand{\Bsrc}{\mathcal B}

\newcommand{\Hgen}{\mathcal H}
\newcommand{\vac}{\lvert 0\rangle}

\newcommand{\ket}[1]{\lvert #1\rangle}
\newcommand{\coef}[1]{\left[#1\right]}
\title{Cyclotomic Resonances and Generic Non-D-finiteness of the Three-Dimensional Ising Model}
\author{Z. C. Tu\\
\small School of Physics and Astronomy, Beijing Normal University, Beijing 100875, China}
\date{Nov. 27, 2026}
\begin{document}
\maketitle
\begin{abstract}
We establish an all-order cyclotomic pole mechanism for the partially resummed high-temperature pressure of the zero-field simple-cubic Ising model. At each principal primitive root of order $j\geq2$ in the squared longitudinal variable, the first pole appears at transverse degree $4j-2$ and is simple under any positive transverse weighting. A connected, transpose-preserving transfer-matrix normal form identifies the pole amplitude with a quadratic sum of minimal $2j$-particle creation amplitudes. On odd-degree trees these amplitudes factor into local three-branch weights. A projective Lorentz--M\"obius algebra gives a determinantal form of the vertex weight, and an unrooted internal-edge product yields a common sign at the principal root, excluding cancellation. The resulting infinitely many first poles prove intrinsic multivariate non-D-finiteness and exclude longitudinal and transverse linear differential equations. As a consequence, we exclude regular Gaussian free-energy representations based on fixed finite-dimensional rational Fourier symbols. A radial obstruction and finite Taylor-minor argument further establish generic one-variable non-D-finiteness along straight high-temperature rays and at fixed longitudinal coupling. With the energy normalization $J_x:J_y:J_z=1:a:b$, we also obtain non-D-finiteness in $t=\tanh(\beta J_x)$ along physical temperature paths for each fixed $a>0$ outside an at most countable set of longitudinal ratios algebraic over $\Q(a)$. The latter conclusion is distinguished from the intrinsic obstruction, since nonalgebraic temperature parametrizations can themselves destroy D-finiteness. A hard-core hopping resolvent describes the highest subsequent poles, with exact finite-graph checks including cycles. Prescribed positive rational ratios, including the completely isotropic point, remain open.
\end{abstract}

\section{Introduction}
The zero-field square-lattice Ising model is the classical example of an exact solution~\cite{onsager,kaufman,baxter}. For the simple cubic lattice no analogous solution is known. Inversion relations and discrete coupling symmetries have provided important constraints on lattice-model solutions, including the cubic model \cite{jaekel1982,jaekel1983}. One approach to the question of solvability is to examine the partially resummed high-temperature expansion, retaining one coupling as a variable while expanding in the other two. Hansel et al.~\cite{hansel} investigated the rational coefficient functions and their denominator structure. Guttmann and Enting~\cite{guttmannenting} proposed a solvability criterion based on the proliferation of coefficient singularities (see also \cite{guttmann2000}). For lattice polygons, Rechnitzer developed a systematic analysis of anisotropic rational generating functions and their denominators~\cite{rechnitzer2003}, and subsequently proved non-D-finiteness for self-avoiding polygons~\cite{rechnitzer}. The implications for the cubic Ising problem are discussed in \cite{viswanathan}. The open issue addressed here is the survival of new denominator factors at arbitrarily high orders, rather than the denominator criterion itself.

Other approaches to solvability and obstructions to
it have different scopes. Istrail's complexity
results concern computing partition functions across
families of nonplanar lattices \cite{istrail};
they do not imply non-D-finiteness of the homogeneous
infinite-volume pressure. Regge and Zecchina
\cite{regge2000} extended the Pfaffian formalism
to lattices of higher topological genus $g$,
expressing the partition function in terms of
$2^{2g}$ Pfaffians associated with different
spin structures. Such a representation need not
reduce, in the thermodynamic limit, to a Gaussian
theory with a fixed finite-dimensional Fourier symbol.

More recently, Niedzi\'{o}\l{}ka and Wojtkiewicz
\cite{niedziolka} identified a sign obstruction to
a conventional local bilinear Grassmann construction
of the cubic Ising high-temperature expansion.
Their obstruction concerns a particular fermionic
representation, whereas our analysis addresses
the differential-algebraic structure of the
thermodynamic pressure directly, without
presupposing a fermionic representation.
As a consequence of our non-D-finiteness theorem,
we also exclude a class of regular Gaussian
free-energy representations with
finite-dimensional rational Fourier symbols~(see Corollary~\ref{cor-gaussian}).

The main obstacle is cancellation. Candidate cyclotomic denominators in individual graphs need not survive the logarithm of the partition function, the sum over insertion orders, or specializations of the couplings. The original inspiration for our approach comes from Bryant's study of Willmore surfaces \cite{bryant1984}: a change of representation, based on Lorentzian conformal geometry, can reveal structures that are hidden in the original variables. In the present problem this idea takes a more elementary form, namely the projective algebra of complex Lorentz boosts.

The proof combines this local algebra with a connected transfer-matrix normal form. The linked-support argument is related in spirit to the linked-cluster property of effective Hamiltonians \cite{bravyi}, but here the transformation preserves transpose symmetry and eliminates nonresonant particle-number grades at a complex root of unity. The lowest connected resonant sources are supported on odd-degree trees. A combinatorial decomposition of insertion times yields a product of local vertex weights, whose determinantal form is obtained through the Lorentz--M\"obius composition law. The symmetric vertex factors give a root-independent edge product and establish a common sign at the principal root. Consequently, the first pole coefficient is a sum of squares of real resonant source amplitudes, with no cancellation. The resulting first-pole theorem establishes intrinsic multivariate non-D-finiteness and provides the basis for generic obstructions along straight high-temperature rays.

A second obstacle concerns fixed microscopic interactions: non-D-finiteness of a multivariate function need not imply that of its one-variable temperature specializations. We address this issue by directional and radial differential-equation obstructions and a specialization argument based on finite Taylor minors. Fixing the energy scale by $J_x=1$ and writing the other interaction strengths as $a,b>0$, we obtain non-D-finiteness along the physical temperature path for each fixed $a>0$ outside a possible countable set of longitudinal ratios algebraic over $\Q(a)$. In particular, for equal in-plane interactions all positive transcendental longitudinal ratios are covered. This physical-path result is a further specialization theorem: for irrational interaction ratios the temperature map is nonalgebraic, so this result alone is not evidence of nonintegrability. The stronger leading-pole resolvent and the direction-budget criterion are established separately in Appendix~\ref{app-higher}; neither is needed for the generic physical-path theorem. We do not claim to rule out all possible closed forms, and we do not settle the fully isotropic point.

\section{Model and principal results}\label{sec-main}
For the zero-field nearest-neighbor model on $\Z^3$, take the ferromagnetic Hamiltonian
\begin{equation}
 \mathcal H_{\rm spin}=-\sum_{i=x,y,z} J_i\sum_{\langle uv\rangle_i}\sigma_u\sigma_v,
 \qquad J_x=1,\quad J_y=a,\quad J_z=b,\quad a,b>0.
\end{equation}
Here $\sigma_u\in\{-1,1\}$, $\langle uv\rangle_i$ ranges over unordered nearest-neighbor bonds parallel to axis $i$, each counted once, and $\beta=1/(k_{\mathrm B}T)$ with $T$ being the temperature. The unit of energy is $J_x$ before normalization. Put $K=\beta J_x$, $K_x=K$, $K_y=aK$, $K_z=bK$, and introduce
\begin{equation}\label{eq-physicalpath}
 t=\tanh K,\quad w(t)=\operatorname{arctanh}t,\quad
 h_c(t)=\tanh(cw(t)),\qquad (x,y,z)=(t,h_a(t),h_b(t)).
\end{equation}
Thus choosing the energy unit does not freeze any spin or restrict the positive interaction ratios. In contrast, setting the high-temperature variable $x$ to one is a singular strong-coupling limit. Setting one interaction identically to zero yields independent square-lattice layers. In the auxiliary multivariate analysis we allow $x,y,z$ to vary independently as complex variables near zero.
We take free boundary conditions on cubic boxes $\Lambda_L=\{-L,\ldots,L\}^3$ and on transverse square boxes $\Omega_L=\{-L,\ldots,L\}^2$. The notation $|A|$ denotes the cardinality of a finite set $A$; in particular $|\Lambda_L|=(2L+1)^3$ and $|\Omega_L|=(2L+1)^2$. Throughout, $\Lambda\nearrow\Z^3$ and $\Omega\nearrow\Z^2$ abbreviate these box limits as $L\to\infty$, not an arbitrary exhaustion. The fraction of sites at the geometric boundary tends to zero in each case. For a finite box $\Lambda$ let $E_i$ be the bond sets in the three directions and $E(\Lambda)=\bigcup_i E_i$. With $\mathcal H_\Lambda$ the restriction of the Hamiltonian to bonds in $\Lambda$, the partition function is $Z_\Lambda=\sum_{\sigma\in\{-1,1\}^{\Lambda}}e^{-\beta\mathcal H_\Lambda(\sigma)}$. Its high-temperature expansion reads
\begin{equation}\label{eq-HT}
 Z_\Lambda=2^{|\Lambda|}\prod_{i=x,y,z}(\cosh K_i)^{|E_i|}
 \mathcal Q_\Lambda,\qquad
 \mathcal Q_\Lambda=\sum_{\substack{F\subseteq E(\Lambda)\\\partial F=\varnothing}}
 x^{|F\cap E_x|}y^{|F\cap E_y|}z^{|F\cap E_z|}.
\end{equation}
Here $\partial F=\{v:\deg_F(v)\text{ is odd}\}$ is the odd-degree boundary of an edge set, not the geometric boundary of a region; $\varnothing$ denotes the empty set. We use $F,G,H$ for edge sets and $S,S'$ for occupied site sets, with $A\triangle B=(A\setminus B)\cup(B\setminus A)$ denoting symmetric difference of sets of the same type. We use the reduced pressure
\begin{equation}\label{eq-phi}
 \Phi(x,y,z)=\lim_{\Lambda\nearrow\Z^3}|\Lambda|^{-1}\log\mathcal Q_\Lambda,
\end{equation}
with the logarithm normalized at zero. Standard results on the thermodynamic limit and the high-temperature cluster expansion \cite{friedli} give its analytic germ (the convergent local Taylor series) at the origin. The alternative cylinder limit used in Appendix~\ref{app-transfer}, first in the longitudinal length and then along the square boxes $\Omega_L$, agrees coefficientwise with this cubic-box definition. The dimensionless physical pressure differs from $\Phi$ by $\log 2+\sum_i\log\cosh K_i$. Define the physical per-site partition function (the exponential of the thermodynamic pressure) by
\begin{equation}\label{eq-Zinf}
 Z_\infty(x,y,z)=\frac{2e^{\Phi(x,y,z)}}{\sqrt{(1-x^2)(1-y^2)(1-z^2)}},
\end{equation}
where the square root is the branch equal to one at the origin.

For a formal or convergent series, $[x^m y^n]f$ denotes extraction of the indicated monomial coefficient, where $m,n$ are nonnegative integer indices. Put $s=z^2$ and write
\begin{equation}\label{eq-partial}
 \Phi(x,y,z)=\sum_{m,n\geq0}r_{mn}(s)x^{2m}y^{2n},\qquad
 H_N(s)=\sum_{m+n=N}r_{mn}(s).
\end{equation}
Independent sublattice gauge transformations give even powers of all three variables in the thermodynamic high-temperature germ. Appendix~\ref{app-transfer} proves coefficientwise rationality $r_{mn}\in\Q(s)$ by stabilization of connected transverse supports. In particular $\Phi(0,0,z)=0$. All uses of these rational coefficients at complex $s$ are coefficientwise; they do not presume convergence of the full transverse sum at a resonance.
Define
\begin{align}
 F_a(t,z)&=\Phi(t,h_a(t),z)=\sum_{N\geq0}H_N^{(a)}(z^2)t^{2N},\label{eq-Fa}\\
 G_{a,b}(t)&=F_a(t,h_b(t)).\label{eq-Gab}
\end{align}
Here $H_N^{(a)}\in\Q(a)(s)$ and $H_N^{(1)}=H_N$. The physical pressure along the path is
\begin{equation}\label{eq-physP}
 P_{a,b}(t)=G_{a,b}(t)+\log 2+\log\cosh w(t)+\log\cosh(aw(t))+\log\cosh(bw(t)).
\end{equation}
A power series is D-finite if its mixed derivatives span a finite-dimensional vector space over the rational function field in its variables \cite{stanley,lipshitz}. In one variable this is equivalent to a nonzero homogeneous linear differential equation with polynomial coefficients. In what follows, algebraicity over a field $k\subset\C$ means membership in its algebraic closure inside $\C$.
The principal notation is summarized in Table~\ref{tab-notation}; the operators and tree quantities appearing there are defined in Sections~\ref{sec-factor} and \ref{sec-resonance} and Appendix~\ref{app-higher}.

\begin{table}[htbp]
\centering\small
\caption{Principal notation used throughout
the paper. The transverse expansion index is
$N=m+n$, corresponding to total degree $2N$
in the transverse high-temperature variables.}\label{tab-notation}
\begin{tabularx}{\linewidth}{@{}lX@{}}\toprule
Symbol & Meaning\\\midrule
$J_x,J_y,J_z;\ a,b$ & Interaction energies; normalized ratios $1:a:b$\\
$K,K_i;\ t,x,y,z;\ s$ & Dimensionless couplings; high-temperature variables; $s=z^2$\\
$\Lambda_L,\Omega_L$ & Finite cubic and transverse square boxes\\
$F,G,H;\ S,S'$ & Edge subsets; occupied vertex subsets\\
$m,n,N;\ j,d$ & Transverse coefficient indices, $N=m+n$; resonance order and $d=2j-1$\\
$\zeta_j,z_j,\delta$ & $e^{2\pi\ii/j}$, $e^{\pi\ii/j}$, and local coordinate $1-s^j$\\
$\Phi,r_{mn},H_N^{(a)}$ & Reduced pressure, rational coefficients, weighted transverse coefficients\\
$g_j, R_j(a)$ & Tree amplitude and normalized first-pole amplitude\\
$\Nop,D,X_e$ & Occupation number, longitudinal propagation, edge flip\\
$\Bsrc_j,\mathsf A_j,p_j(m,n)$ & Resonant source, hard-core hopping matrix, pole order\\\bottomrule
\end{tabularx}
\end{table}

\begin{theorem}[First cyclotomic poles]\label{thm-first}
For $j\geq2$, let $d=2j-1$ and $\zeta_j=e^{2\pi\ii/j}$. Every $r_{mn}(s)$ with $m+n<d$ is regular at $\zeta_j$; at $m+n=d$ the poles are at most simple. For every $a>0$, $H_N^{(a)}$ is regular there for $N<d$, whereas
\begin{equation}\label{eq-positive}
 R_j(a):=\lim_{s\to\zeta_j}(1-s^j)H_d^{(a)}(s)
 =\sum_{m+n=d} C_j(m,n)a^{2n}>0,
 \quad C_j(m,n)=\lim_{s\to\zeta_j}(1-s^j)r_{mn}(s).
\end{equation}
All $C_j(m,n)$ are nonnegative real numbers. We write $R_j=R_j(1)$. The ordinary residue of $H_d^{(a)}$ is $-R_j(a)/(j\zeta_j^{j-1})$.
\end{theorem}

\begin{theorem}[Intrinsic and directional non-D-finiteness]\label{thm-main}
For every $a>0$, $F_a(t,z)$ and $e^{F_a(t,z)}$ admit no nonzero rational-coefficient linear differential equation involving only $\partial_z$, or only $\partial_t$. Consequently $\Phi(x,y,z)$, $\Phi(t,t,z)$, their exponentials and the corresponding physical pressures and physical per-site partition functions are non-D-finite as multivariate germs.
\end{theorem}

\begin{theorem}[Generic specializations in high-temperature variables]\label{thm-special}
The following assertions hold.
\begin{enumerate}
\item For fixed $a>0$, the formal transverse germ $F_a(t,z_0)$ is non-D-finite in $t$ for all $0<z_0<1$ outside an at most countable set algebraic over $\Q(a)$. This is an analytic-germ assertion wherever the transverse expansion converges.
\item For fixed $a>0$, $F_a(t,bt)$ is non-D-finite for all $b>0$ outside an at most countable set algebraic over $\Q(a)$. In particular this covers the straight high-temperature rays of the equal-transverse model when $a=1$.
\end{enumerate}
The conclusions in (1) and (2) also hold for the reduced per-site partition function obtained by exponentiating $F_a$.
For each fixed $a>0$, the same conclusions hold with $F_a(t,z)$ replaced by $\Phi(t,at,z)$ and with its exponential. In particular, the straight rays $(x,y,z)=(t,at,bt)$ involve only algebraic substitutions in the original variables.
\end{theorem}

\begin{theorem}[Generic non-D-finiteness on physical temperature paths]\label{thm-physical}
Fix any $a>0$. There exists an at most countable set $\mathcal E_a\subset\overline{\Q(a)}$ such that for every $b>0$ outside $\mathcal E_a$, all four analytic germs
\begin{equation}\label{eq-four}
 G_{a,b}(t),\qquad e^{G_{a,b}(t)},\qquad P_{a,b}(t),\qquad e^{P_{a,b}(t)}
\end{equation}
are non-D-finite in $t$. In particular for $a=1$, every positive transcendental $b$ is covered. For each $a>0$ the allowed $b$ have full Lebesgue measure in every positive interval.
\end{theorem}

Theorems~\ref{thm-special} and \ref{thm-physical} describe possible exceptional sets, not the behavior of any prescribed positive algebraic parameter. In particular, the completely isotropic specialization is not settled. Non-D-finiteness after a nonalgebraic change of temperature variable is not by itself a criterion of nonintegrability.

The proofs proceed as follows. Appendix~\ref{app-transfer} establishes rationality and coefficientwise stabilization. Sections~\ref{sec-factor} and \ref{sec-resonance}, with detailed proofs in Appendices~\ref{app-cones} and \ref{app-normal}, provide the local amplitudes and the connected normal form used in the proof of Theorem~\ref{thm-first} in Section~\ref{sec-equal}. Section~\ref{sec-dfinite} proves Theorems~\ref{thm-main} and \ref{thm-special} and derives the Gaussian representation obstruction in Corollary~\ref{cor-gaussian}; Section~\ref{sec-physical} proves Theorem~\ref{thm-physical}. The higher-pole results are proved separately in Appendices~\ref{app-higher}--\ref{app-embedding}. Figure~\ref{fig-proofmap} in Appendix~\ref{app-roadmap} summarizes these dependencies.

\section{Local Lorentz--M\"obius algebra and resonant tree factorization}\label{sec-factor}
For $j\geq2$, let $\mathcal T$ be a tree on $2j$ vertices with every degree equal to one or three, and write $V(\mathcal T)$ and $E(\mathcal T)$ for its vertex and edge sets. Such a tree has $d=2j-1$ edges and $j-1$ trivalent vertices. For its edge subsets write $k(F)=|\partial F|/2$. At a primitive $j$th root $\zeta$ define the square-free creation recursion
\begin{align}
 q(\varnothing)&=1,\\
 q(F)&=\frac{\zeta^{k(F)}}{1-\zeta^{k(F)}}\sum_{G\subsetneq F}q(G),
 \qquad \varnothing\ne F\subsetneq E(\mathcal T),\label{eq-rec}\\
 g_j(\mathcal T;\zeta)&=\sum_{G\subsetneq E(\mathcal T)}q(G).\label{eq-g}
\end{align}
The boundary map on subsets of tree edges is injective; the boundary of the full edge set is the full vertex set. Thus $1\leq k(F)\leq j-1$ for every nonempty proper subset, and all denominators in \eqref{eq-rec} are nonzero. The recursion will be identified with the transfer-matrix creation coefficient in Appendix \ref{app-normal}.

Choose a leaf as the root. At each trivalent vertex, let the two outward subtrees have $2m-1$ and $2n-1$ vertices. These numbers are odd: cutting a tree edge leaves components in which exactly one original odd degree is reduced by one. Define
\begin{equation}\label{eq-vertex}
 \mathcal V_\zeta(m,n)=
 \frac{2\zeta^{m+n-1}(1-\zeta)}{(1-\zeta^m)(1-\zeta^n)(1-\zeta^{m+n-1})}.
\end{equation}
\begin{theorem}[Tree amplitude factorization]\label{thm-tree}
For every such tree and every primitive $j$th root $\zeta$,
\begin{equation}\label{eq-treeprod}
 g_j(\mathcal T;\zeta)=\prod_{v:\deg v=3}\mathcal V_\zeta(m_v,n_v).
\end{equation}
The expression is independent of the root. Equivalently, if deleting $v$ leaves three branches with sizes $2m_v-1,2n_v-1,2r_v-1$, its local factor is
\begin{equation}\label{eq-symvertex}
 \mathcal V_\zeta(m_v,n_v,r_v)=
 -\frac{2(1-\zeta)}{(1-\zeta^{m_v})(1-\zeta^{n_v})(1-\zeta^{r_v})},\qquad m_v+n_v+r_v=j+1.
\end{equation}
\end{theorem}

\begin{proof}[Proof idea]
Expanding \eqref{eq-rec} yields a sum over ordered partitions of the tree edges into nonempty blocks. Each block represents one insertion time; the gaps between consecutive blocks are resummed by geometric series. At resonance the total time weight is invariant under common time translation. Fixing a root edge time at zero makes the two child-edge time differences at every branching vertex independent integer coordinates. The time-dependent weight then separates into local three-branch weights. A rational sum over the six ordering cones and their boundaries evaluates each local weight as \eqref{eq-vertex}. Appendix \ref{app-cones} gives the complete rational-summation argument, including the treatment of nonconvergent geometric series at unit roots. Equation \eqref{eq-symvertex} follows from $\zeta^j=1$ and $m+n-1=j-r$.
\end{proof}

\paragraph{Projective Lorentz--M\"obius algebra.}
The local factor \eqref{eq-vertex} admits a two-dimensional matrix representation that makes the geometric motivation of our approach concrete. In this representation, the nonlinear composition of the Cayley coordinates is induced by ordinary multiplication of complex Lorentz matrices. To exhibit this structure, write $\zeta=e^{-2\eta}$ and introduce
\begin{equation}\label{eq-mobius}
 u_k=\frac{1+\zeta^k}{1-\zeta^k}=\coth(k\eta),\qquad
 \mathsf L(\eta)=\begin{pmatrix}\cosh\eta&\sinh\eta\\\sinh\eta&\cosh\eta\end{pmatrix}.
\end{equation}
The matrix $\mathsf L(\eta)$ has determinant one and preserves the bilinear form $\operatorname{diag}(1,-1)$ under transpose. For complex $\eta$ this is a complexified Lorentz boost, not a real spacetime symmetry of the Ising model. Set
\begin{equation}
 \mathsf M_k=\begin{pmatrix}u_k&1\\1&u_k\end{pmatrix},\qquad
 \mathsf L(k\eta)=\sinh(k\eta)\mathsf M_k.
\end{equation}
The matrices $\mathsf M_k$ are representatives of the projective transformations. For $k,l,k+l$ nonzero modulo $j$, the multiplication law $\mathsf L(k\eta)\mathsf L(l\eta)=\mathsf L((k+l)\eta)$ gives
\begin{equation}
 u_{k+l}=\frac{u_ku_l+1}{u_k+u_l},\qquad u_{j-k}=-u_k.
\end{equation}
More directly, the product $\mathsf L(m\eta)\mathsf L(n\eta)\mathsf L(-\eta)$ gives the shifted composition law
\begin{equation}\label{eq-shiftedmob}
 u_{m+n-1}=\frac{(u_mu_n+1)u_1-u_m-u_n}{(u_m+u_n)u_1-u_mu_n-1}.
\end{equation}
All coordinates here are finite for the branch indices in Theorem~\ref{thm-tree}. The denominator in \eqref{eq-shiftedmob} is the off-diagonal entry of $\mathsf M_m\mathsf M_n\operatorname{adj}(\mathsf M_1)$ and is nonzero because $1\leq m+n-1\leq j-1$; $\operatorname{adj}$ is the classical adjugate.

The local ordering-cone calculation in Appendix~\ref{app-cones} yields, in Cayley coordinates,
\begin{equation}\label{eq-mobver}
 2\mathcal V_\zeta(m,n)=u_{m+n-1}(u_m+u_n)-u_mu_n-1.
\end{equation}
Using \eqref{eq-shiftedmob}, put $U=u_m+u_n$ and $P=u_mu_n+1$. The numerator after substitution is $P^2-U^2=(u_m^2-1)(u_n^2-1)$. This proves the local determinant identity
\begin{align}
 \mathcal V_\zeta(m,n)
 &=\frac{(u_m^2-1)(u_n^2-1)}{2[(u_m+u_n)u_1-u_mu_n-1]}\\
 &=\frac{\det(\mathsf M_m\mathsf M_n)}{2(\mathsf M_m\mathsf M_n\operatorname{adj}(\mathsf M_1))_{12}}.\label{eq-detvertex}
\end{align}
Equivalently, the denominator is $2\sinh((m+n-1)\eta)/(\sinh(m\eta)\sinh(n\eta)\sinh\eta)$, so this determinant expression is
\begin{equation}\label{eq-hypvertex}
 \mathcal V_\zeta(m,n)=\frac{\sinh\eta}{2\sinh(m\eta)\sinh(n\eta)\sinh((m+n-1)\eta)}.
\end{equation}
Thus the matrix composition evaluates the vertex weight in local variables; it does not replace the combinatorial proof of the ordering-cone factorization. At the principal root, $\eta=-\pi\ii/j$, the boost obeys $\mathsf L(j\eta)=-I_2$ ($I_2$ denotes the two-dimensional identity), and its projective class has order $j$. This identifies cyclotomic resonance with projective periodicity of the local propagation parameter.

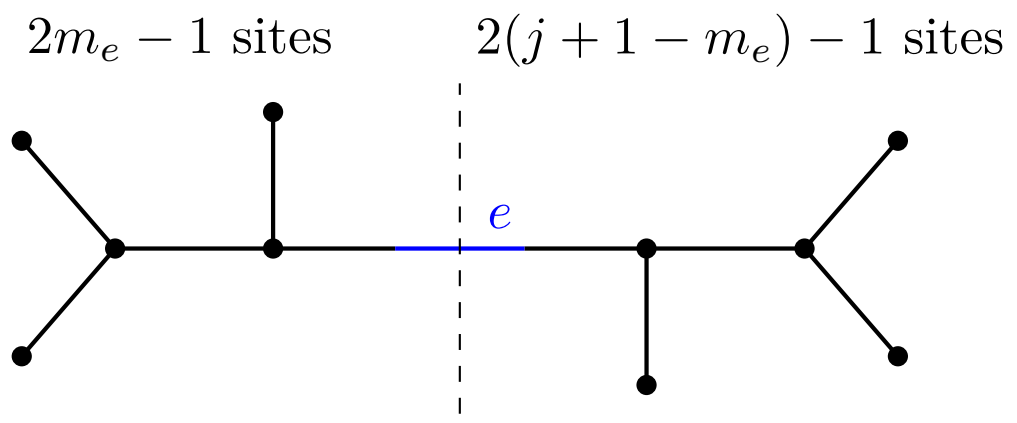
\begin{figure}[htbp]
\centering
\begin{tikzpicture}[scale=0.95, every node/.style={font=\small}]
\draw[thick] (-2.4,0)--(-1.3,0)--(-0.45,0) (0.45,0)--(1.3,0)--(2.4,0);
\draw[thick] (-2.4,0)--(-3.05,0.75) (-2.4,0)--(-3.05,-0.75) (-1.3,0)--(-1.3,0.95) (1.3,0)--(1.3,-0.95) (2.4,0)--(3.05,0.75) (2.4,0)--(3.05,-0.75);
\draw[thick,blue] (-0.45,0)--(0.45,0); \draw[dashed] (0,-1.15)--(0,1.15);
\foreach \x/\y in {-2.4/0,-1.3/0,1.3/0,2.4/0,-3.05/.75,-3.05/-.75,-1.3/.95,1.3/-.95,3.05/.75,3.05/-.75} {\fill (\x,\y) circle (2pt);}
\node[blue,above right] at (0.05,0) {$e$};
\node at (-1.95,1.45) {$2m_e-1$ sites};
\node at (1.95,1.45) {$2(j+1-m_e)-1$ sites};
\end{tikzpicture}
\caption{An internal edge of an odd-degree tree and its complementary branches. The drawing shows $j=5$ and $m_e=3$; the split sizes, rather than a chosen root, enter \eqref{eq-edgeprod}.}\label{fig-split}
\end{figure}

\begin{corollary}[Unrooted internal-edge product]\label{cor:edges}
Let $E_{\rm int}(\mathcal T)$ consist of edges joining two trivalent vertices. There are $j+1$ leaves and $j-2$ internal edges. On deleting an internal edge $e$, let its two components have sizes $2m_e-1$ and $2(j+1-m_e)-1$. Then
\begin{equation}\label{eq-edgeprod}
 g_j(\mathcal T;\zeta)=\frac{(-2)^{j-1}}{(1-\zeta)^2
 \displaystyle\prod_{e\in E_{\rm int}(\mathcal T)}(1-\zeta^{m_e})(1-\zeta^{j+1-m_e})}.
\end{equation}
At the principal root, writing $\theta=\pi/j$,
\begin{equation}\label{eq-edgesine}
 g_j(\mathcal T;\zeta_j)=\frac{(-1)^{j-1}}{2^{j-1}\sin^2\theta
 \displaystyle\prod_{e\in E_{\rm int}(\mathcal T)}\sin(m_e\theta)\sin((m_e-1)\theta)}.
\end{equation}
Empty products are one. Both expressions are invariant under reversing the choice of component of any edge.
\end{corollary}
\begin{proof}
The degree sum of a tree gives $j+1$ leaves, $j-1$ trivalent vertices and hence $j-2$ internal edges. Multiply the symmetric factors \eqref{eq-symvertex}. Each leaf contributes a denominator $1-\zeta$; each internal edge supplies the two denominators corresponding to its complementary components. The numerator contributes $(1-\zeta)^{j-1}$, leaving the factor $(1-\zeta)^2$ in the denominator of \eqref{eq-edgeprod}. For \eqref{eq-edgesine}, apply $1-e^{2\ii k\theta}=-2\ii e^{\ii k\theta}\sin(k\theta)$ to every factor, using $\sin((j+1-m_e)\theta)=\sin((m_e-1)\theta)$. There are $2j-2$ factors and the sum of their indices is $2+(j-2)(j+1)=j(j-1)$. Thus their phase is $(-\ii)^{2j-2}e^{\ii\pi(j-1)}=1$, which gives the displayed sign from $(-2)^{j-1}$. Equivalently the sign follows by multiplying \eqref{eq-hypvertex}. For the internal edges $2\leq m_e\leq j-1$, so no factor vanishes.
\end{proof}
The unordered multiset of internal-edge split sizes determines the amplitude; trees with identical split multisets have identical amplitudes, although the converse is not asserted. The complementary components associated with an internal edge are illustrated in Fig.~\ref{fig-split}.

\begin{corollary}[Common sign and noncancellation]\label{cor:sign}
At the principal root $\zeta_j=e^{2\pi\ii/j}$, for every odd-degree tree on $2j$ vertices,
\begin{equation}\label{eq-sign}
 (-1)^{j-1}g_j(\mathcal T;\zeta_j)>0.
\end{equation}
At every primitive root, the amplitude of an individual tree is nonzero.
\end{corollary}
\begin{proof}
Equation~\eqref{eq-edgesine} proves the assertion directly. At the vertex level, with $\theta=\pi/j$, Eq.~\eqref{eq-vertex} becomes
\begin{equation}
 \mathcal V_{\zeta_j}(m,n)=
 -\frac{\sin\theta}{2\sin(m\theta)\sin(n\theta)\sin((m+n-1)\theta)}<0,
\end{equation}
since all three integer arguments are between $1$ and $j-1$. Multiply the $j-1$ factors. Nonvanishing at any primitive root follows directly from \eqref{eq-vertex}.
\end{proof}
The common sign is asserted at the principal root only. Other roots need not have this sign property; nonvanishing of a rational coefficient at them will follow from Galois conjugacy.

For a tree with a chain of $j-1$ trivalent vertices, the factors in \eqref{eq-treeprod} have arguments $(1,k)$, $1\leq k<j$. Using the standard cyclotomic product gives the particularly simple value
\begin{equation}\label{eq-chain}
 g_j^{\rm chain}=\prod_{k=1}^{j-1}\frac{2\zeta^k}{(1-\zeta^k)^2}
 =(-1)^{j-1}\frac{2^{j-1}}{j^2}.
\end{equation}
The product formula also reduces the computation for a specified tree from subset enumeration to $O(j)$ cyclotomic-field arithmetic operations after tabulating the powers of $\zeta$ and branch sizes. This counts arithmetic operations in the cyclotomic field, not bit operations. It does not provide a polynomial algorithm for the full lattice coefficient, which still involves lattice embeddings and paths.

\section{Connected normal form and first resonant sources}\label{sec-resonance}
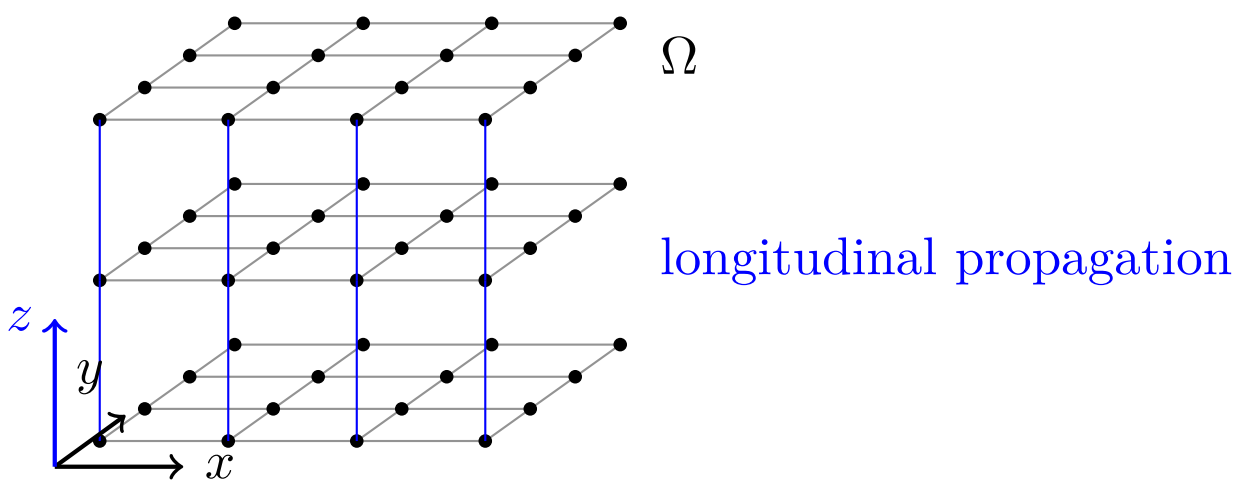
\begin{figure}[htbp]
\centering
\begin{tikzpicture}[scale=.85,every node/.style={font=\small}]
\foreach \k in {0,1,2}{
 \begin{scope}[yshift=\k*1.25cm]
 \foreach \i in {0,1,2,3}{\draw[gray!85] (\i,0)--({\i+1.05},.75); \draw[gray!85] ({.35*\i},{.25*\i})--({3+.35*\i},{.25*\i});}
 \foreach \i in {0,1,2,3}{\foreach \h in {0,1,2,3}{\fill ({\i+.35*\h},{.25*\h}) circle(1.5pt);}}
 \end{scope}}
 \foreach \i in {0,1,2,3}{\foreach \k in {0,1}{\draw[blue,thin] (\i,{1.25*\k})--(\i,{1.25*(\k+1)});}}
 \draw[->,thick] (-.35,-.2)--(.65,-.2) node[right] {$x$};
 \draw[->,thick] (-.35,-.2)--(.2,.2) node[above left] {$y$};
 \draw[->,thick,blue] (-.35,-.2)--(-.35,.95) node[left] {$z$};
 \node[right] at (4.2,3.0) {$\Omega$};
 \node[right,blue] at (4.2,1.4) {longitudinal propagation};
\end{tikzpicture}
\caption{Transverse square layers and the longitudinal transfer direction. The labels $x,y,z$ denote the corresponding bond weights. The transfer matrix propagates along the blue direction.}\label{fig-layers}
\end{figure}

\paragraph{Coefficientwise convention.} All formal transverse expansions below are taken at a fixed finite target degree before passing to the transverse thermodynamic limit. After introducing the local coordinate $\delta=1-s^j$ and scaled variables $X,Y$, a statement $O(\delta^q)$ means that each relevant coefficient in $X,Y$ is divisible by $\delta^q$ in the ring of holomorphic germs at $\delta=0$. No bound uniform in transverse degree or volume, or convergence for fixed complex $X,Y$, is implied. Formal inverses are taken in the transverse-adic ring when their constant matrices are invertible.
Let $\Omega$ be one of the finite transverse square boxes (Fig.~\ref{fig-layers}). Its character-basis state space is $\mathcal H_\Omega=\bigotimes_{v\in\Omega}\C^2$, with basis $\ket{S}$ indexed by subsets $S\subseteq\Omega$. Equivalently the character of a spin configuration indexed by $S$ is $\prod_{v\in S}\sigma_v$, with normalization in the orthonormal character basis. The local occupation operator obeys $\widehat n_v\ket{S}=\mathbf1_{\{v\in S\}}\ket{S}$, and
\begin{equation}
 \Nop=\sum_{v\in\Omega}\widehat n_v,\qquad
 D\ket{S}=z^{|S|}\ket{S},\qquad
 X_e\ket{S}=\ket{S\triangle\{u,v\}}\quad(e=\{u,v\}).
\end{equation}
Here $D=z^{\Nop}$ is the normalized longitudinal propagation matrix and $\vac=\ket{\varnothing}$ is the vacuum. The particles are binary character occupations, not additional physical particles. Since $X_e$ preserves parity of the occupation number, we work in the even sector. The superscript $\mathsf T$ on matrices and vectors denotes ordinary transpose, not complex conjugation.

A symmetric transfer matrix, up to scalar factors analytic in $x,y$, is
\begin{equation}\label{eq-transfer}
 \mathsf T=D^{1/2}e^V D^{1/2},\qquad V=\sum_{e\subset\Omega}\operatorname{arctanh}(t_e)X_e,
\end{equation}
where we first keep all edge variables independent, then set $t_e=x$ or $y$. Its vacuum branch has eigenvalue $1$ at zero transverse variables. The log of this eigenvalue gives the reduced pressure after subtraction of the elementary transverse $\log\cosh$ factors. Details and the thermodynamic coefficient limit are in Appendix \ref{app-transfer}.

Fix $j\geq2$ and a primitive root $\zeta$, choosing a local branch of $z=\sqrt{s}$ near $s=\zeta$. A monomial operator of particle-number grade $k$ satisfies $[\Nop,A_k]=2kA_k$. Nonresonant grades $j\nmid k$ can be removed to any finite order by an orthogonal (transpose-preserving, not unitary) formal similarity transform. At the order being removed, its generator is
\begin{equation}\label{eq-homological}
 S_k=-\frac{A_k}{z^k-z^{-k}},\qquad S_{-k}=-S_k^{\mathsf T}.
\end{equation}
The denominator is nonzero near $\zeta$ for these grades. Because nested commutators of operators of disjoint supports vanish, connected transverse support is preserved. Appendix \ref{app-normal} makes these statements precise.

The resulting resonant logarithm has no connected vacuum-to-$2j$-particle source below order $d=2j-1$: at least $2j$ vertices are required, and a connected graph with that many vertices has at least $2j-1$ edges. At equality every vertex is flipped an odd number of times, all edges occur once, and the graph is a tree with degrees one or three. The coefficient of each such tree is $z^{-j}g_j(\mathcal T;\zeta)$ at resonance, as proved by Lemma~\ref{lem-matching} in Appendix~\ref{app-normal}. The factor $z^{-j}=\pm1$ is common to all trees and reflects the symmetric transfer-matrix convention. The same lowest-degree bound holds for the vacuum off-diagonal block of the exponential: read a product of positive-degree resonant logarithm factors from right to left acting on the vacuum. Before the first creation, a grade-zero factor cannot leave the vacuum; the first factor leaving it has positive grade $mj$ and hence degree at least $2mj-1\geq d$. At degree $d$ there can be only this one positive-degree factor, necessarily with $m=1$. The external diagonal factors do not change this counting. Disjoint pairs of flips can produce $2j$ particles in fewer than $d$ insertions in the original transfer matrix; such disconnected contributions are not independent resonant sources in the connected logarithmic normal form. The distinction is essential to the pole threshold.

For a set $S\subseteq\Omega$ of $2j$ occupied sites, define the homogeneous polynomial
\begin{equation}\label{eq-B}
 \Bsrc_{j,S}(X,Y)=\sum_{\substack{\mathcal T\subset\Omega\ \text{odd-degree tree}\\ V(\mathcal T)=S}}
 z^{-j}g_j(\mathcal T;\zeta)X^{n_x(\mathcal T)}Y^{n_y(\mathcal T)},
\end{equation}
where $z^2=\zeta$ on the right, $n_x(\mathcal T)$ and $n_y(\mathcal T)$ count its horizontal and vertical edges, and $X,Y$ are auxiliary transverse weights. The sum counts edge sets, not choices of roots or embeddings of a labelled abstract tree. Let $\mathsf A_x,\mathsf A_y$ be the adjacency matrices for moving one of $2j$ hard-core particles to an unoccupied nearest-neighbor site in the two transverse directions. The leading resonant hopping operator is
\begin{equation}
 \mathsf A_j(X,Y)=X\mathsf A_x+Y\mathsf A_y.
\end{equation}

\section{First cyclotomic poles at arbitrary transverse ratios}\label{sec-equal}
For the first poles we need only the lowest off-diagonal source, not a resummation of hopping.
\begin{proof}[Proof of Theorem~\ref{thm-first}]
Fix a finite layer and work through total transverse degree $2d$ in the connected normal form, where $d=2j-1$. In the block resonant with the vacuum, write $\mathsf T'$ in vacuum/complement blocks as
\begin{equation}
 \mathsf T'_{\rm res}=\begin{pmatrix}v&B^{\mathsf T}\\ B&C\end{pmatrix},\qquad
 \lambda=v+B^{\mathsf T}(\lambda\Id-C)^{-1}B.
\end{equation}
These identities are formal in the transverse variables for generic $s$; the truncated removal of other sectors suffices at the indicated degrees. The coefficients of $v$ and of $B$ are holomorphic at $s=\zeta_j$. The source $B$ starts at degree $d$, and its degree-$d$ part is supported only in the $2j$-particle sector. The constant transverse term of $\lambda\Id-C$ is $1-s^{mj}$ in the $2mj$-particle sector. Consequently the Schur correction starts at degree $2d$ and there uses only the inverse constant gap $1/(1-s^j)$. In particular, all smaller-degree logarithmic coefficients are holomorphic. Since $v$ has constant term one, taking the logarithm leaves this first singular term unchanged. With $\Bsrc_{j,S}$ as in \eqref{eq-B}, the homogeneous degree-$2d$ pole polynomial is therefore
\begin{equation}\label{eq-firstpoly}
 \sum_{m+n=d} C_j(m,n)X^{2m}Y^{2n}
 =\lim_{\Omega\nearrow\Z^2}\frac{1}{|\Omega|}
 \sum_{\substack{S\subseteq\Omega\\|S|=2j}}\Bsrc_{j,S}(X,Y)^2.
\end{equation}
The limit is coefficientwise: all trees have a fixed number of edges, and for a given occupied set only finitely many tree edge sets occur. The stabilized coefficient limit follows from Appendix~\ref{app-transfer}. Every coefficient of the right side is nonnegative at the principal root by Corollary~\ref{cor:sign}; odd direction powers vanish by the sublattice symmetries.

Since $h_a(t)=at+O(t^3)$, the coefficient of $t^{2d}$ in $F_a$ equals the substitution $(X,Y)=(1,a)$ into the degree-$2d$ part of $\Phi$, plus linear combinations of smaller-degree coefficients that are holomorphic at the resonance. Thus
\begin{equation}\label{eq-Rj}
 R_j(a)=\lim_{\Omega\nearrow\Z^2}\frac{1}{|\Omega|}
 \sum_{\substack{S\subseteq\Omega\\|S|=2j}}\Bsrc_{j,S}(1,a)^2.
\end{equation}
For a horizontal chain of $j-1$ trivalent core vertices, attach one vertical leaf per core vertex and the two horizontal end leaves (for $j=2$, a three-edge star). This is a valid embedded tree with $n_x=j$ and $n_y=j-1$. Its translates alone, using \eqref{eq-chain} and common-sign noncancellation, yield the explicit lower bound
\begin{equation}\label{eq-lower}
 R_j(a)\geq \frac{4^{j-1}}{j^4}a^{2j-2}>0\qquad(a>0).
\end{equation}
This proves the regularity, simplicity bound and strict positivity in Theorem~\ref{thm-first}. The residue formula follows from $(1-s^j)'|_{s=\zeta_j}=-j\zeta_j^{j-1}$.
\end{proof}
For $a=1$, rationality also implies that every primitive root of order $j$ is a pole of $H_d$, by Galois conjugation. The principal root suffices for our differential-equation arguments; for arbitrary fixed $a$ we make no sign claim at other roots.

For example, $j=2$ corresponds to a three-edge star. Its amplitude at $s=-1$ is $-1/2$ in the convention of \eqref{eq-g}. There are four such stars per lattice site (the center with any three of its four neighbors), and each occupied set determines its star uniquely. Equation \eqref{eq-Rj} at $a=1$ gives
\begin{equation}\label{eq-star}
 \lim_{s\to-1}(1-s^2)H_3(s)=4(1/2)^2=1.
\end{equation}
In particular, the two directional amplitudes at $(m,n)=(1,2),(2,1)$ are both $1/2$. For comparison, Hansel et al. \cite{hansel} proposed, based on their series, the expression
\begin{equation}\label{eq-hansel}
 r_{21}(s)=\frac{1+54s+293s^2+472s^3+293s^4+54s^5+s^6}{(1-s)^5(1+s)}.
\end{equation}
Its predicted normalized pole amplitude is $\lim_{s\to-1}(1-s^2)r_{21}(s)=1/2$, in agreement with our all-order mechanism. We use this comparison as a normalization check, not as a proof of the full rational expression \eqref{eq-hansel}.

For $j=3,4$ the trivalent cores are paths and the amplitude is given by \eqref{eq-chain}. Enumerating square-lattice edge sets modulo translations, then grouping by occupied set in \eqref{eq-firstpoly}, gives the exact rational predictions in Table~\ref{tab-amplitudes}. Reflections and rotations are retained as distinct when they are not translations. For general $j$, it is the common sign rather than a classification of embeddings or occupied sets that ensures that \eqref{eq-Rj} cannot vanish.

\begin{table}[htbp]
\centering
\caption{First normalized pole amplitudes from lattice tree enumeration. Entries related by $m\leftrightarrow n$ are shown once.}\label{tab-amplitudes}
\begin{tabular}{rrrrll}\toprule
$j$ & $d$ & Trees & Occupied sets & $R_j$ & $(m,n):C_j(m,n)$\\\midrule
2 & 3 & 4 & 4 & $1$ & $(1,2):1/2$\\
3 & 5 & 18 & 18 & $32/9$ & $(1,4):16/81,\ (2,3):128/81$\\
4 & 7 & 76 & 76 & $19$ & $(2,5):3/2,\ (3,4):8$\\\bottomrule
\end{tabular}
\end{table}

For instance $R_2(a)=(a^2+a^4)/2$ and $R_3(a)=[16(a^2+a^8)+128(a^4+a^6)]/81$. The direction-budget theorem of Appendix~\ref{app-higher} also gives the precise small-$a$ order $R_j(a)\asymp a^{2\lceil(j-1)/2\rceil}$ as $a\downarrow0$ at each fixed $j$ (the symbol $\asymp$ means bounded above and below by positive constant multiples for sufficiently small $a$). This is a coefficientwise statement, not a uniform estimate in $j$.

Appendix~\ref{app-check} provides complementary computational checks. Table~\ref{tab-treecheck} tests the tree identities by finite-field subset recursion; Table~\ref{tab-finitetree} compares the predicted first poles with a direct finite-tree transfer-matrix recursion. Table~\ref{tab-gridcheck} extends the latter comparison to a transverse graph containing cycles, including several highest-pole coefficients beyond the first resonance. These finite-order checks support, but do not replace, the analytic proofs.

\section{Directional obstructions and generic specialization}\label{sec-dfinite}
We record three elementary consequences of the first-pole mechanism. They separate the model-specific problem of proving noncancellation from the differential-algebraic problem of taking specializations.
\begin{lemma}[Finite-pole criterion]\label{lem-pole}
Let $F(u,z)=\sum_{n\geq0}f_n(z)u^n$ with $f_n\in\C(z)$. If $F$ satisfies a nonzero equation in $\partial_z$ with coefficients in $\C(u,z)$, then the union of the finite poles of the rational functions $f_n$ is a finite set.
\end{lemma}
\begin{proof}
Clear denominators, remove the maximal common power of $u$, and write the equation as $\sum_{k=0}^m u^k L_k(z,\partial_z)F=0$, with $L_0\ne0$ and all coefficients polynomial. Let $a(z)$ be the leading coefficient of $L_0$ in derivative order. If $z_*$ is the first pole at index $n_*$ and $a(z_*)\ne0$, coefficient comparison gives
\begin{equation}
 L_0f_{n_*}=-\sum_{k=1}^{\min(m,n_*)}L_k f_{n_*-k},
\end{equation}
whose right side is holomorphic at $z_*$. The highest derivative in $L_0$ raises the pole order of $f_{n_*}$ strictly above the orders of all lower derivative terms (the order-zero case is immediate). Thus its pole cannot disappear, a contradiction. Every first pole is therefore among the finitely many zeros of $a$.
\end{proof}

\begin{lemma}[Transverse obstruction]\label{lem-transverse}
Let $k=\Q(a)$ for a fixed real $a$, and let $A(u,s)=\sum_{N\geq0}f_N(s)u^N$, $f_N\in k(s)$. Suppose $\zeta_j$ is a pole first appearing at $N=2j-1$ for every $j\geq2$. Then $A$ satisfies no nonzero linear equation in $\partial_u$ over $\C(u,s)$.
\end{lemma}
\begin{proof}
If an equation exists over $\C(u,s)$, clear denominators and bound the degrees and derivative orders of its polynomial coefficients. Equating coefficients of $u$ and clearing the rational denominators in $s$ gives a system of homogeneous linear equations over $k$ on a finite number of unknown scalar coefficients. A nontrivial solution over $\C$ implies one over $k$: a finite subset of rows spans the row space, and its rank is unchanged by scalar extension. The D-finite/P-recursive equivalence \cite{stanley}, or direct coefficient comparison, then gives a nontrivial recurrence
\begin{equation}\label{eq-prec}
 \sum_{\ell=0}^{M}P_\ell(N,s)f_{N+\ell}(s)=0,
 \quad P_\ell\in k[N,s],\quad P_M\ne0,
\end{equation}
for all sufficiently large $N$. Substituting $N=2j-1-M$, the unique potentially singular term forces $P_M(2j-1-M,\zeta_j)=0$. Along primes $j$, the degree of $\zeta_j$ over $k$ tends to infinity: for algebraic $a$ it is at least $(j-1)/[k:\Q]$, and for transcendental $a$ the pure transcendental extension leaves the cyclotomic degree unchanged. For sufficiently large such primes, this degree exceeds the fixed $s$-degree of $P_M$. Thus $P_M(2j-1-M,s)$ is identically zero for infinitely many $j$, which contradicts $P_M\ne0$.
\end{proof}

\begin{lemma}[Finite Taylor minors]\label{lem-generic}
Let $A(t,b)=\sum_{N\geq0}c_N(b)t^N$ with $c_N\in k(b)$, where $k\subset\C$ is a field of characteristic zero. Suppose there is no nonzero differential equation in $\partial_t$ over $\C(t,b)$. Then there is an at most countable exceptional set contained in $\overline{k}$ outside which $A(t,b_0)$ is well defined coefficientwise and is non-D-finite over $\C(t)$.
\end{lemma}
\begin{proof}
For each $r,D_{\max}\geq0$, the $(r+1)(D_{\max}+1)$ series $t^q\partial_t^\ell A$, $0\leq q\leq D_{\max}$, $0\leq\ell\leq r$, are linearly independent over $k(b)$ by the hypothesis. Some finite square Taylor-coefficient minor of these columns is a nonzero rational function $\Delta_{r,D_{\max}}(b)\in k(b)$. Remove its zeros and poles and the poles of all coefficients $c_N$. Each removed set is finite and algebraic over $k$, so their union is at most countable. At any remaining point the minor has full rank over $\C$ and excludes every equation with the specified degree and order bounds. Taking the union over all bounds proves the assertion. No uniform degree or order bound for a hypothetical specialized equation is required. The converse is not claimed: a vanishing chosen minor need not imply D-finiteness.
\end{proof}

\begin{proof}[Proof of Theorems~\ref{thm-main} and \ref{thm-special}]
Apply Lemma~\ref{lem-pole} to $F_a$, using the infinitely many points $z_j=e^{\pi\ii/j}$ of Theorem~\ref{thm-first}. This excludes all longitudinal equations. Lemma~\ref{lem-transverse}, applied with $u=t^2$ and $s=z^2$, excludes the transverse equations: if the even series were D-finite in $t$ over $\C(z)$, algebraic substitution $t=\sqrt u$ would give D-finiteness in $u$ over $\C(z)$, and descent along the finite extension $\C(s)\subset\C(z)$ would give one over $\C(s)$. These are standard algebraic-substitution and scalar-restriction properties of linear differential equations. Applying Lemma~\ref{lem-generic} with parameter $z$ then gives the fixed-$z_0$ assertion. Appendix~\ref{app-transfer} shows that the rational coefficient denominators of $F_a$ have no roots with $0<z_0<1$ (all arise from $1-s^k$). The formal statement thus concerns well-defined coefficients throughout this real interval. It implies an analytic statement at least near the origin, by high-temperature analyticity; no assertion of convergence for the entire interval is required here.

For the straight high-temperature paths $z=bt$, a hypothetical equation with polynomial coefficients in $t,b$, written in $t\partial_t$, pulls back via $b=z/t$ to an equation in the operator $t\partial_t+z\partial_z$ with Laurent-polynomial coefficients in $t$ and polynomial coefficients in $z$. After clearing negative powers of $t$, the identity holds for small nonzero $t$ and $z/t$ in a bounded open set. Both sides extend holomorphically to a bidisc about the origin; the identity theorem therefore permits coefficient comparison in $t$ with $z$ independent. Remove the common lowest $t$-power. Its constant-in-$t$ coefficients are polynomials $p_\ell(z)$, not all zero. On the coefficient of $t^{2d}$ this operator acts as $2d+z\partial_z$, whereas all terms from smaller coefficients are holomorphic at $z_j$. If $r$ is the highest index with $p_r\ne0$, then for infinitely many $z_j$ the coefficient $p_r(z_j)z_j^r$ is nonzero, and the highest derivative gives an uncancellable pole of order $r+1$. This contradicts the equation. The Taylor coefficients of $F_a(t,bt)$ are in $k[b]$, hence Lemma~\ref{lem-generic} gives the asserted generic straight-path obstruction. This argument does not assume that coefficient poles survive the full infinite sum at a fixed path.

The same arguments work for $e^{F_a}$, since $H_0^{(a)}=0$ and each coefficient of its exponential is the corresponding coefficient of $F_a$ plus a polynomial in smaller coefficients. Its first pole at every $z_j$ is therefore unchanged. In the fully anisotropic case a putative longitudinal equation can be divided by the maximal common power of $x-y$ in its polynomial coefficients and restricted to $x=y=t$, giving a nonzero forbidden equation. The longitudinal obstruction for the physical pressures and physical per-site partition functions in the multivariate high-temperature variables follows by adding the elementary logarithmic terms or multiplying by the algebraic factors in \eqref{eq-Zinf}, respectively (and using their inverses). For the purely algebraic rays, replace $h_a(t)$ throughout by $at$. The first-pole polynomial \eqref{eq-firstpoly} gives exactly the same $R_j(a)$, and all transverse coefficients still belong to $\Q(a)(s)$. The longitudinal, transverse, radial and Taylor-minor arguments therefore apply unchanged. No general closure of non-D-finite functions under exponentiation has been assumed.
\end{proof}

An additional consequence of
Theorem~\ref{thm-main} concerns Gaussian
free-energy representations. The obstruction
is independent of the number of auxiliary
fermionic degrees of freedom, provided that
the resulting Fourier symbols have fixed
finite dimension and satisfy the rationality
and regularity conditions specified below.

\begin{corollary}
[Obstruction to regular rational-symbol
Gaussian free energies]
\label{cor-gaussian}

Let $\boldsymbol w=(x,y,z)$ denote the
independent high-temperature variables.
The reduced pressure $\Phi(\boldsymbol w)$
of the anisotropic simple-cubic Ising model
does not admit a representation of the form
\begin{equation}
\label{eq-gaussian-pressure}
\Phi(\boldsymbol w)
={}\Phi_0(\boldsymbol w)+\sum_{\nu=1}^{M}
\frac{c_\nu}{(2\pi\ii)^{d_\nu}}
\oint_{\mathbb T^{d_\nu}}
\log\frac{
 D_\nu(\boldsymbol w;\boldsymbol q)}
 {D_\nu(\boldsymbol 0;\boldsymbol q)}
\prod_{\mu=1}^{d_\nu}
\frac{\dd q_\mu}{q_\mu},
\end{equation}
under the following assumptions:

\begin{enumerate}
\item
$M$ and $d_\nu$ are fixed positive integers,
and $c_\nu\in\C$ are constants independent
of the coupling variables.

\item
$\Phi_0(\boldsymbol w)$ is a D-finite
analytic germ at $\boldsymbol w=\boldsymbol 0$.

\item
For each $\nu$,
\[
D_\nu(\boldsymbol w;\boldsymbol q)
\in\C(x,y,z,q_1,\ldots,q_{d_\nu})
\]
is a rational function, where
$\boldsymbol q=(q_1,\ldots,q_{d_\nu})$.

\item
The integration cycle is the fixed,
positively oriented product torus
\[
\mathbb T^{d_\nu}
=
\{\boldsymbol q\in(\C^\times)^{d_\nu}:
 |q_\mu|=1,\ 1\leq\mu\leq d_\nu\}.
\]
There exist a polydisc $U$ about
$\boldsymbol w=\boldsymbol 0$ and an open
neighborhood $V_\nu$ of $\mathbb T^{d_\nu}$
such that $D_\nu$ is holomorphic and
nonvanishing on $U\times V_\nu$.
The logarithm in
\eqref{eq-gaussian-pressure} is the
analytic branch normalized to vanish
at $\boldsymbol w=\boldsymbol 0$.
\end{enumerate}

In particular, the excluded class includes
regular Gaussian free-energy representations
whose functions $D_\nu$ are determinants of
fixed finite-dimensional Fourier symbols:
\begin{equation}
\label{eq-gaussian-symbol}
D_\nu(\boldsymbol w;\boldsymbol q)
=
\det\widehat{\mathsf A}_\nu
(\boldsymbol q;\boldsymbol w),
\end{equation}
where
\begin{equation}
\label{eq-gaussian-fourier}
\widehat{\mathsf A}_\nu
(\boldsymbol q;\boldsymbol w)
=
\sum_{\boldsymbol\ell\in\mathcal R_\nu}
\mathsf A_{\nu,\boldsymbol\ell}
(\boldsymbol w)
\boldsymbol q^{\boldsymbol\ell},
\end{equation}
$\mathcal R_\nu\subset\Z^{d_\nu}$ is finite,
\[
\boldsymbol q^{\boldsymbol\ell}
=\prod_{\mu=1}^{d_\nu}q_\mu^{\ell_\mu},
\]
and the coefficient matrices have fixed finite
dimension and entries in $\C(x,y,z)$.

\end{corollary}

\begin{proof}
Suppose that a representation of the form
\eqref{eq-gaussian-pressure} exists.
Write
\[
I_\nu(\boldsymbol w)
=
\frac{1}{(2\pi\ii)^{d_\nu}}
\oint_{\mathbb T^{d_\nu}}
\log\frac{
 D_\nu(\boldsymbol w;\boldsymbol q)}
 {D_\nu(\boldsymbol 0;\boldsymbol q)}
\prod_{\mu=1}^{d_\nu}
\frac{\dd q_\mu}{q_\mu}.
\]

The regularity assumption and compactness
of the integration torus ensure that,
after shrinking $U$ if necessary, the
normalized logarithm is well defined
and differentiation under the integral
sign is justified.

For each coupling variable
$w_i\in\{x,y,z\}$, we obtain
\begin{equation}
\label{eq-gaussian-derivative}
\partial_{w_i}I_\nu
=
\frac{1}{(2\pi\ii)^{d_\nu}}
\oint_{\mathbb T^{d_\nu}}
\frac{\partial_{w_i}D_\nu}{D_\nu}
\prod_{\mu=1}^{d_\nu}
\frac{\dd q_\mu}{q_\mu}.
\end{equation}
Write $D_\nu=P_\nu/Q_\nu$ with coprime
polynomials in the coupling and integration variables.
The holomorphic, nonvanishing hypothesis implies that
neither $P_\nu$ nor $Q_\nu$ vanishes on $U\times V_\nu$:
otherwise an irreducible factor of one polynomial would
have a zero hypersurface meeting this open set, and
at a generic point of that hypersurface the other
polynomial would not vanish, contradicting the
holomorphic unit property of $D_\nu$.

Put $k=\C(x,y,z)$ and define the localized coordinate ring
\[
\mathcal S_\nu=
 k[q_1^{\pm1},\ldots,q_{d_\nu}^{\pm1},(P_\nu Q_\nu)^{-1}].
\]
Logarithmic differentiation gives
\[
 \frac{\partial_{w_i}D_\nu}{D_\nu}
 =\frac{\partial_{w_i}P_\nu}{P_\nu}
 -\frac{\partial_{w_i}Q_\nu}{Q_\nu}.
\]
Thus the integrand in \eqref{eq-gaussian-derivative}
(including the differential form $\prod_\mu\dd q_\mu/q_\mu$),
and all its mixed coupling derivatives, belong to the
space of top-degree relative algebraic differential
forms $\Omega^{d_\nu}_{\mathcal S_\nu/k}$. Parameter
differentiation can increase the pole orders, but cannot
introduce new polar divisors in the integration variables
over $k$.

The affine variety $\operatorname{Spec}\mathcal S_\nu$
is a smooth open subset of the algebraic torus over
the characteristic-zero field $k$. By the finiteness
theorem for algebraic de Rham cohomology
\cite{grothendieck1966,lairez2016}, the quotient
\[
 H^{d_\nu}_{\mathrm{dR}}(\operatorname{Spec}\mathcal S_\nu/k)
 =\frac{\Omega^{d_\nu}_{\mathcal S_\nu/k}}
 {\dd_{\boldsymbol q}\Omega^{d_\nu-1}_{\mathcal S_\nu/k}}
\]
is finite-dimensional over $k$. Here $\dd_{\boldsymbol q}$
acts only in the integration variables; the top-degree
quotient computes de Rham cohomology since the variety
is smooth and affine of dimension $d_\nu$.

Consequently, for each $i$ a finite subset of the
classes of the mixed derivatives spans all their classes
over $k$. Each resulting linear relation between forms
holds modulo $\dd_{\boldsymbol q}\eta$, for a globally
regular relative algebraic $(d_\nu-1)$-form $\eta$ on
$\operatorname{Spec}\mathcal S_\nu$. For parameter values
in $U$ outside the zero sets of the finitely many
parameter denominators of a given relation and of
$\eta$, the specialized forms are defined in a
neighborhood of the fixed torus. The integral of the
exact term is zero by Stokes' theorem, because the torus
is closed. After clearing the parameter denominators,
the period relation extends to all of $U$ by the
holomorphic identity theorem. This reasoning applies
to each relation separately, so all of them hold for
the analytic germs at the high-temperature origin.
Consequently, the mixed derivatives of each
$\partial_{w_i}I_\nu$ span a finite-dimensional
$k$-vector space, and each such first derivative
is D-finite.
Every mixed derivative of $I_\nu$
of positive total order is a derivative
of at least one of these three
D-finite first derivatives.
Adjoining $I_\nu$ itself therefore gives
a finite-dimensional derivative space
over $\C(x,y,z)$.
Thus $I_\nu$ is D-finite in the
multivariate sense.
Since $\Phi_0$ is D-finite by assumption,
and the D-finite class is closed under
finite linear combinations
\cite{lipshitz}, the right-hand side
of \eqref{eq-gaussian-pressure}
is D-finite.
This contradicts
Theorem~\ref{thm-main}.

Finally, under the assumptions of
\eqref{eq-gaussian-fourier}, every
matrix entry is a Laurent polynomial
in $\boldsymbol q$ with rational
coefficients in $\boldsymbol w$.
Its determinant is therefore rational
in all variables, and the preceding
argument applies whenever the stated
regularity condition holds.
\end{proof}

The scope of Corollary~\ref{cor-gaussian}
is specifically the thermodynamic
free-energy representation
\eqref{eq-gaussian-pressure}.
It does not exclude finite-volume Pfaffian
representations whose number of terms or
Fourier-symbol dimension increases with
system size, such as topological
constructions of the type considered
by Regge and Zecchina \cite{regge2000}.
Nor does it exclude nonlocal or
interacting fermionic representations
that fall outside the stated assumptions.
In particular, the corollary does not
settle the existence of a Gaussian
representation restricted to the
completely isotropic model.

\section{Fixed physical interactions and the temperature path}\label{sec-physical}
The straight path $z=bt$ used above is generally not a fixed-interaction temperature path. We now exploit the energy normalization to obtain the physical version, with $z=h_b(t)$. The proof uses coordinates in which this path is exactly a line.

\subsection*{Radial obstruction in physical coupling coordinates}
Set $K=w(t)$, $L=w(z)$, and define
\begin{equation}\label{eq-KLF}
 \mathcal F_a(K,L)=\Phi(\tanh K,\tanh(aK),\tanh L),
 \qquad L=bK.
\end{equation}
Let $\mathcal U_a(K,L)$ stand for any of the four functions
\begin{equation}\label{eq-Ufour}
 \mathcal F_a,\quad e^{\mathcal F_a},\quad
 \mathcal P_a:=\mathcal F_a+\log2+\log\cosh K+\log\cosh(aK)+\log\cosh L,\quad e^{\mathcal P_a}.
\end{equation}
Expand $\mathcal U_a(K,L)=\sum_{N\geq0}U_N(L)K^N$. Put
\begin{equation}\label{eq-Lj}
 z_j=e^{\pi\ii/j},\qquad L_j=w(z_j)=\tfrac12\log\cot\tfrac{\pi}{2j}+\tfrac{\pi\ii}{4},\qquad N_j=4j-2.
\end{equation}
The $L_j$ are distinct, nonzero, and lie away from zeros of $\cosh L$. All coefficients $U_N$ for $N<N_j$ are holomorphic at $L_j$, and $U_{N_j}$ has a simple pole there. Indeed $\tanh K=K+O(K^3)$ and $\tanh L$ has a nonzero derivative at $L_j$, so the first-pole property follows from Theorem~\ref{thm-first}. Exponentiating the reduced pressure preserves the first pole as before. Adding the elementary pressure terms is holomorphic at $L_j$. For the full per-site partition function the factor $2\cosh K\cosh(aK)\cosh L$ multiplies the first-pole coefficient by $2\cosh L_j\ne0$ and adds only lower regular contributions. In particular the argument for the physical partition function does not require $h_a$ or $h_b$ to be algebraic in $t$.

These coefficients admit meromorphic continuation in $L$ except for the elementary logarithm in $U_0$ of $\mathcal P_a$. For definiteness all coefficient comparisons and continuations may be made in the strip $|\operatorname{Im}L|<\pi/2$, where $\cosh L$ has no zero and $\log\cosh L$ has an analytic branch normalized at zero. Every $L_j$ belongs to this strip. The continuation is of individual coefficients only.

\begin{proposition}[No uniform physical-path equation]\label{prop-radial}
For each $a>0$ and each $\mathcal U_a$ in \eqref{eq-Ufour}, the germ $A(t,b)=\mathcal U_a(w(t),bw(t))$ satisfies no nonzero linear equation in $\partial_t$ with coefficients in $\C(t,b)$.
\end{proposition}
\begin{proof}
Suppose such an equation exists. Clear denominators and express it using the Euler derivative $t\partial_t$ to obtain
\begin{equation}\label{eq-uniform}
 \sum_{\ell=0}^r P_\ell(t,b)(t\partial_t)^\ell A(t,b)=0,
 \qquad P_\ell\in\C[t,b],\quad (P_0,\ldots,P_r)\ne0.
\end{equation}
On the path $t=\tanh K$, $L=bK$, the Euler derivative pulls back to
\begin{equation}\label{eq-radial}
 \mathcal D=c(K)(K\partial_K+L\partial_L),\qquad
 c(K)=\frac{\sinh K\cosh K}{K}=1+O(K^2).
\end{equation}
Substitute $b=L/K$ in \eqref{eq-uniform} and multiply by a sufficiently large power of $K$ to clear its denominators. The resulting coefficient functions of $K,L$ are holomorphic near $(0,0)$, and polynomial in $L$ with coefficients holomorphic in $K$. Indeed, the original formal identity is an analytic identity for sufficiently small $t$ and $b$ in a bounded open parameter neighborhood: the high-temperature germ and the maps $w,h_b$ are jointly analytic there. The map $(K,b)\mapsto(K,L=bK)$ has nonzero Jacobian $K$ for $K\ne0$, so its image contains a nonempty open subset in a sufficiently small connected bidisc of analyticity about $(0,0)$. After clearing denominators, the multivariable identity theorem gives the resulting identity on that bidisc. This justifies the substitution without evaluating $L/K$ at $K=0$. Divide out the maximal common power of $K$ from these coefficient functions. At least one has a nonzero constant-in-$K$ coefficient. This gives an equation of the form
\begin{equation}\label{eq-radialleading}
 \sum_{\ell=0}^r [p_\ell(L)+O(K)]\mathcal D^\ell\mathcal U_a(K,L)=0,
 \qquad p_\ell\in\C[L],\quad (p_0,\ldots,p_r)\ne0.
\end{equation}
The $O(K)$ terms here are analytic in $K$ with polynomial-in-$L$ Taylor coefficients. The operator $\mathcal D$ preserves the $K$-adic filtration; on degree $N$ its leading action is $N+L\partial_L$. Extract the coefficient of $K^{N_j}$ in \eqref{eq-radialleading} and continue it to a neighborhood of $L_j$. We obtain
\begin{equation}\label{eq-contradiction}
 \sum_{\ell=0}^r p_\ell(L)(N_j+L\partial_L)^\ell U_{N_j}(L)=V_j(L),
\end{equation}
where $V_j$ depends only on smaller-degree coefficients and their derivatives, so is holomorphic at $L_j$. Let $q$ be the highest index with $p_q\ne0$. The coefficient of the highest derivative is $p_q(L)L^q$. Since the $L_j$ are distinct and nonzero, this coefficient is nonzero at all but finitely many $L_j$. For such $j$, the simple pole of $U_{N_j}$ produces a pole of order $q+1$ on the left, which no lower derivative can cancel. This contradicts \eqref{eq-contradiction}, including when $q=0$.
\end{proof}

\subsection*{From the uniform obstruction to fixed ratios}
\begin{proof}[Proof of Theorem~\ref{thm-physical}]
The Taylor coefficients of each $A(t,b)$ are polynomials in $b$ over $k=\Q(a)$, except for the harmless constant $\log 2$ in the full pressure. This follows from the high-temperature graph expansion and from $h_b(t)=bt+(b-b^3)t^3/3+\cdots$, whose coefficients are rational polynomials in $b$ at every order. For the full pressure use $\mathcal P_a-\log2$; adding a constant preserves D-finiteness, so Proposition~\ref{prop-radial} also excludes its uniform equation (otherwise adjoining the constant gives an equation for $\mathcal P_a$). Lemma~\ref{lem-generic} therefore applies to all four functions. The finite union of the corresponding countable exceptional sets proves Theorem~\ref{thm-physical}. For each finite real $a,b$ the temperature-path series is an analytic germ by the high-temperature expansion, since $(t,h_a(t),h_b(t))$ tends to the origin as $t\to0$.
\end{proof}

\begin{remark}
The exceptional set is a possible set of failures of the Taylor-minor certificates, not an assertion that any given algebraic ratio is D-finite. It can include every positive rational number. The theorem concerns the physically standard high-temperature variable $t$, and it is not a claim that D-finiteness is invariant under arbitrary analytic changes of temperature coordinate.
\end{remark}

\section{Discussion and conclusion}\label{sec-discussion}
The local odd-degree-tree factorization is the model-specific ingredient of our proof. Its common sign eliminates cancellation at the principal primitive root, and the connected transpose-preserving normal form converts it into a positive quadratic first-pole coefficient. The differential-equation obstruction then has a wider scope: infinitely many new coefficient poles exclude a unified longitudinal, transverse or radial linear equation under the conditions above. A finite Taylor-minor argument transfers these obstructions to generic one-variable paths, without identifying the actual singularities of the resummed function on any such path. The normalization $J_x=1$ retains all positive physical interaction ratios, and Theorem~\ref{thm-physical} concerns temperature variation at fixed Hamiltonian, rather than simply independent complex variables.

The distinction between intrinsic and parametrization-dependent non-D-finiteness is essential. For instance, a single four-spin rectangular loop has reduced pressure $\phi(x,z)=\tfrac14\log(1+x^2z^2)$, a D-finite germ. For irrational $b$, the continuation of $h_b(t)$ has poles at the infinitely many distinct finite points
\begin{equation}
 t_k=\ii\tan\left(\frac{(2k+1)\pi}{2b}\right),\qquad k\in\Z.
\end{equation}
Indeed these are images under $t=\tanh K$ of poles of $\tanh(bK)$. They are nonzero, finite, and distinct for irrational $b$, and are reached by continuing the germ along paths avoiding the isolated singularities in the $K$ plane and the branch points of $\operatorname{arctanh}t$. At each such point a continued branch of $\phi(t,h_b(t))$ has a logarithmic singularity, since $1+t^2h_b(t)^2$ has a pole of order two there. A D-finite germ has only finitely many possible finite singular points under continuation, namely zeros of a leading polynomial coefficient. Thus even this elementary finite model becomes non-D-finite under the irrational physical parametrization. By contrast, Theorem~\ref{thm-main} in the original variables and Theorem~\ref{thm-special} on the purely algebraic rays are not subject to this ambiguity: D-finiteness of the original multivariate germ would imply D-finiteness of its regular linear specializations. The physical-path theorem is accordingly a further specialization result, not a standalone criterion for integrability.

For equal transverse interactions, the $b=0$ endpoint is the D-finite square-lattice reduced pressure $G_{1,0}(t)=\Phi_{\rm 2D}(t,t)$, as follows from the exact solution \cite{onsager,guttmann2000}. By Theorem~\ref{thm-physical}, $G_{1,b}(t)$ is non-D-finite for almost every $b>0$, including ratios arbitrarily close to zero. The physical pressure itself is continuous at dimensional reduction: writing $p$ for the dimensionless thermodynamic pressure, the finite-volume inequality $|\partial p_\Lambda/\partial K_z|\leq1$ gives
\begin{equation}\label{eq-continuity}
 |p(K_x,K_y,K_z)-p(K_x,K_y,0)|\leq |K_z|.
\end{equation}
The change in differential-algebraic behavior is not a thermodynamic discontinuity. When $z=0$, the longitudinal operator $D=z^{\Nop}$ collapses to the vacuum projector, and \eqref{eq-Tnorm} has rank one with nonzero eigenvalue $\sum_{\partial F=\varnothing}\prod_{e\in F}t_e$, the reduced partition function of a single layer. All cyclotomic denominators become unity under this specialization. When $x=0$ or $y=0$, the transverse graph instead becomes a union of chains, which cannot support odd-degree trees on more than two vertices. This geometric observation is consistent with the known $s=1$ denominator structure of the square-lattice coefficients \cite{onsager,hansel,guttmann2000}; it alone does not exclude all higher-order resonances. For irrational $a$, the parametrization of the two-dimensional model by $t,h_a(t)$ is not in general an algebraic substitution, so the asserted D-finite endpoint above is restricted to $a=1$ (and algebraic substitutions where applicable).

Corollary~\ref{cor-gaussian} connects the
intrinsic non-D-finiteness result with
the conventional free-fermion approach
to exact solvability. It shows that increasing the number of
auxiliary fermionic variables to an
arbitrarily large but fixed finite number
cannot overcome the obstruction, as long
as the thermodynamic pressure retains
a regular logarithmic torus-integral
representation with rational Fourier
symbols. This leaves open more general constructions,
including representations with
system-size-dependent topological sectors
and genuinely interacting fermionic
descriptions.

It is also instructive to distinguish energy normalization from fixing a high-temperature variable to one. For $K_x,K_y,K_z\geq0$ the all-aligned contribution and an upper bound by decoupled $x$-chains give
\begin{equation}
 K_x+K_y+K_z\leq p(K_x,K_y,K_z)\leq\log(2\cosh K_x)+K_y+K_z.
\end{equation}
Hence for fixed nonnegative $y,z<1$,
\begin{equation}\label{eq-frozen}
 \lim_{x\uparrow1}\Phi(x,y,z)=\log(1+y)+\log(1+z).
\end{equation}
The reduced per-site partition function has the limit $(1+y)(1+z)$. This is a boundary value in the real ferromagnetic region, not a specialization of the high-temperature analytic germ: the even Taylor coefficient $[y^2]\Phi(x,y,0)=x^2/(1-x^2)$ already diverges there. Thus the strong-coupling limit freezes degrees of freedom and cannot represent a mere choice of energy units.

The fully isotropic case remains the principal unresolved specialization. For $a=1$ define the actual set $\mathcal S=\{b\geq0:G_{1,b}\text{ is D-finite in }t\}$. We have $0\in\mathcal S$ and $\mathcal S\setminus\{0\}\subseteq\overline{\Q}\cap\mathbb R_{>0}$, but cannot exclude $1\in\mathcal S$. The natural conjecture is $\mathcal S=\{0\}$. A particularly important intermediate goal is to exclude a prescribed positive rational ratio $b=p/q$, since in that case
\begin{equation}
 \left(\frac{1+h_b(t)}{1-h_b(t)}\right)^q
 =\left(\frac{1+t}{1-t}\right)^p,
\end{equation}
so the parametrization is algebraic. For example, $b=2$ gives the concrete problem of deciding D-finiteness of $\Phi(t,t,2t/(1+t^2))$. No proof at this ratio is given here, and no claim is made that it is easier than the isotropic case. The countable exceptional set may contain every positive rational number, so the generic result alone does not exclude any of these specific values. A possible route is to construct arithmetic certificates for the Taylor minors at prescribed algebraic $b$, especially $b=1$, or to control the resummation of multiple cyclotomic sectors on that path. Neither approach is accomplished here. In particular, the highest-pole expansion near $1-t^{2j}=0$ on the isotropic line involves the effective variable $t/(1-t^{2j})$, which is unbounded there. A coefficientwise pole theorem does not by itself control that regime or establish a natural boundary of the full pressure.

The internal-edge formula \eqref{eq-edgeprod} suggests a complementary local-algebra approach to the quantitative amplitudes. One may group trees by their edge-split multisets and seek extremal amplitudes or a gluing recursion that retains the split weights as trees grow. The leading-pole resolvent and sharp directional budget in Appendix~\ref{app-higher} already retain information beyond the first-pole obstruction. At lower pole orders, however, repeated edges and cycles become relevant. This raises the question of whether a higher-dimensional geometric representation, in the spirit of Bryant's original construction, could extend the local projective composition law \eqref{eq-shiftedmob} to such processes and to interacting resonance sectors. Such an extension would require new identities; neither a bulk Lorentz symmetry nor a closed renormalization scheme for the cubic pressure is established here.

The point $s=1$, where all even sectors are resonant, also requires a separate analysis. Determining the complete lowest denominator, bounding bit complexity of the full lattice sums, and deciding the algebraic exceptional parameters remain open problems. Our results exclude D-finite representations in the specified variables, not nonlinear differential equations or every possible notion of a closed-form solution.

\appendix
\section{Transfer matrix, rationality and thermodynamic coefficients}\label{app-transfer}
For a finite transverse box $\Omega$, the one-site longitudinal matrix in the spin basis has entries $e^{K_z\sigma\sigma'}$. Its two eigenvalues are $2\cosh K_z$ and $2\sinh K_z$. Dividing by the first and making the character-basis change gives $\mathrm{diag}(1,z)$. Tensor products produce $D=z^{\Nop}$. Multiplication by $e^{K_e\sigma_r\sigma_s}$ in the spin basis becomes $e^{K_e X_e}$ in the character basis. Thus the normalized transfer matrix is similar to \eqref{eq-transfer}, and also to $De^V$. Its vacuum eigenvalue $\lambda_\Omega$ is uniquely determined as a formal series in the independent transverse $t_e$ with constant term one, for $s$ away from the finitely many resonances of the finite layer. Working in the even-particle sector, the diagonal unperturbed eigenvalue on $\ket{S}$ is $s^{|S|/2}$. It is convenient to remove the transverse scalar factors by using instead
\begin{equation}\label{eq-Tnorm}
 \widetilde{\mathsf T}=D^{1/2}\prod_e(\Id+t_eX_e)D^{1/2}.
\end{equation}
Since $X_e^2=\Id$ and the $X_e$ commute, \eqref{eq-Tnorm} differs from \eqref{eq-transfer} only by $\prod_e(1-t_e^2)^{1/2}$. The two forms give identical singular Laurent coefficients at all positive transverse orders, because the difference in their logarithmic eigenvalues is analytic in the transverse variables and independent of $s$.

In the nonsymmetric form $D\prod_e(\Id+t_eX_e)$ of \eqref{eq-Tnorm}, normalize the vacuum eigenvector by its vacuum component one. Comparing each transverse monomial in the eigenvalue equation, the nonvacuum components are recursively obtained by division by $1-s^{|S|/2}$. The vacuum component gives the eigenvalue coefficients without any additional division. Hence the log-eigenvalue coefficients are rational functions of $s$ over $\Q$. This is a rational identity, initially obtained for small $s$. It does not use an eigenvalue ordering for complex $s$.

To take the transverse thermodynamic limit, consider a monomial in edge variables with positive exponents on a finite set of edges $E'$. Set all other edge variables to zero. The layer Hilbert space factorizes over the connected components of $(\Omega,E')$. Both $D$ and the transverse transfer matrix factorize, and the distinguished vacuum eigenvalue factorizes as well. Its logarithm is therefore additive. Any monomial using edges in two different connected components has zero coefficient. For total transverse degree $M$, the nonzero monomials consequently have connected support containing at most $M+1$ vertices. After setting the couplings uniform in each direction, the coefficient per site is a finite sum of translation classes of such monomials, with boundary errors vanishing as $\Omega\nearrow\Z^2$. This proves rationality and stabilization of the thermodynamic coefficients of the log-eigenvalue.

For real couplings sufficiently close to zero and a fixed layer, the vacuum branch gives the pressure of the infinite longitudinal cylinder by the positive transfer matrix; its high-temperature Taylor coefficients agree with those of the usual graph expansion. In the subsequent layer limit these stabilized coefficients agree with \eqref{eq-phi}: for each fixed total three-dimensional Taylor order the connected graph expansion involves only finite supports. Equivalently, the usual high-temperature absolute convergence permits taking the box limits for the analytic germ. The extension of each stabilized coefficient to rational $s$ is then unique. No large-volume convergence near $s=\zeta$ is asserted or required.

\section{Proof of the local tree identity}\label{app-cones}
\subsection{Ordered blocks and relative times}
Expanding \eqref{eq-rec} recursively gives
\begin{equation}\label{eq-blocks}
 g_j(\mathcal T;\zeta)=\sum_{(B_1,\ldots,B_\ell)}\prod_{r=1}^{\ell-1}
 \frac{\zeta^{k(F_r)}}{1-\zeta^{k(F_r)}},\qquad F_r=B_1\cup\cdots\cup B_r,
\end{equation}
where the sum runs over all ordered partitions of $E(\mathcal T)$ into nonempty blocks. Each factor in \eqref{eq-blocks} is the rational value of a geometric sum over a positive integer time gap. At a vertex of degree three, sort the three incident insertion times as $t_1\leq t_2\leq t_3$ and define $\tau_v=t_1-t_2+t_3$; at a leaf let $\tau_v$ be its sole insertion time. For any time configuration the total number of occupied-site time steps before the last insertion at $t_{\max}$ is
\begin{equation}
 \sum_v(t_{\max}-\tau_v).
\end{equation}
This follows by following the occupation bit at each vertex through its one or three flips. The interval between the $r$th and $(r+1)$st blocks has $|\partial F_r|=2k(F_r)$ occupied sites. Thus the product of gap weights in \eqref{eq-blocks} is precisely the rational sum over times of $z^{2jt_{\max}-\sum_v\tau_v}$ with $z^2=\zeta$. At resonance $z^{2j}=1$, this weight becomes $z^{-\sum_v\tau_v}$ and is invariant under common integer time translation. Fix the time of the root edge at zero. Each ordered partition with positive time gaps has exactly one such time configuration, and conversely.

At each trivalent vertex use the two differences between its child-edge times and its parent-edge time as independent integer variables $u,v$. These variables at all trivalent vertices parametrize every time configuration uniquely. Indeed the graph of parent-to-child edges is a tree, so there is no cycle consistency condition. If the two outward branches have $2m-1$ and $2n-1$ vertices, translating the branch times by $u$ or $v$ changes their cumulative $\tau$ by $(2m-1)u$ or $(2n-1)v$, respectively. It follows either by induction down the tree or by counting the vertices below each edge that the weight $z^{-\sum_v\tau_v}$ factors into a product of local weights
\begin{equation}\label{eq-localweight}
 z^{-(2m-1)u-(2n-1)v-\tau(0,u,v)},\qquad
 \tau(0,u,v)=\min(0,u,v)-\operatorname{med}(0,u,v)+\max(0,u,v).
\end{equation}
All leaves below a child edge are counted in its branch size. The root leaf has time zero.

\subsection{The six ordering cones}
Set $Q=z^{-2}$, $L_k=Q^k/(1-Q^k)$ and let $A=L_m$, $B=L_n$, $C=L_{m+n-1}$. The elementary identity $L_{-k}=-1-L_k$ is useful. The six strict ordering regions for $(u,v)$ and their rational sums of \eqref{eq-localweight} are listed in Table~\ref{tab-cones}. The boundary rays and the origin are treated separately below.
\begin{table}[htbp]
\centering
\caption{Rational sums of the local weight over the six strict ordering cones. Here $A=L_m$, $B=L_n$, $C=L_{m+n-1}$; the equal-time strata are not included.}\label{tab-cones}
\begin{tabular}{ll@{\hspace{2em}}ll}
\toprule
Region & Sum & Region & Sum\\\midrule
$0<u<v$ & $CB$ & $0<v<u$ & $CA$\\
$u<0<v$ & $L_{-m}B$ & $v<0<u$ & $AL_{-n}$\\
$u<v<0$ & $L_{-(m+n-1)}L_{-m}$ & $v<u<0$ & $L_{-(m+n-1)}L_{-n}$\\\bottomrule
\end{tabular}
\end{table}
For instance, in the first region $u=a$, $v=a+b$ with $a,b\geq1$, so the weight is $Q^{(m+n-1)a+nb}$. The other rows follow by the corresponding change of positive gap variables. On each of the lines $u=0$, $v=0$ and $u=v$, the two nonzero rays sum to $-1$, and the origin contributes $1$. The total is therefore
\begin{equation}\label{eq-localcalc}
 2\big[C(1+A+B)-AB\big]
 =\frac{2Q^{m+n-1}(1-Q)}{(1-Q^m)(1-Q^n)(1-Q^{m+n-1})}.
\end{equation}
The last expression is unchanged by $Q\mapsto Q^{-1}$ and is \eqref{eq-vertex} when $Q=\zeta^{-1}$.

\subsection{Justification of rational factorization at resonance}
We use the following elementary rational-summation lemma, a special case of the standard generating-function valuation for rational cones \cite{barvinok}.
\begin{lemma}[Rational cone subdivision]\label{lem-cones}
Let $L\simeq\Z^N$ be an integer lattice and let $C\subset L\otimes\mathbb R$ be a pointed, relatively open rational polyhedral cone whose closure is pointed. Suppose $C$ is partitioned into finitely many relatively open rational cones $C_i$ (including lower-dimensional strata), and let $\chi:L\to\C^\times$ be a character. For auxiliary characters $w^v=\prod_{h=1}^Nw_h^{v_h}$ define the weighted integer-point generating functions by rational continuation of the sums $\sum_{v\in C\cap L}\chi(v)w^v$ and similarly for $C_i$. Then
\begin{equation}
 G_C(w)=\sum_i G_{C_i}(w)
\end{equation}
as rational functions. Products of cones in independent lattice coordinates give products of generating functions. If all the displayed rational expressions are regular at $w=(1,\ldots,1)$, these identities remain true there.
\end{lemma}
\begin{proof}
Choose a real linear functional $\ell$ strictly positive on all nonzero vectors of the closed pointed cone. On an open domain with $\log|w|+\log|\chi|$ close to $-\ell$, the series converge absolutely (the integer-point count in balls grows at most polynomially). The subdivision identity follows there by disjointness, and the product identity follows by absolute convergence in the independent coordinates. Rationality follows by a rational simplicial subdivision, using fundamental parallelepipeds in the appropriate sublattices and geometric sums along primitive generators, with boundary strata treated separately. Alternatively, for every cone used below the positive integer gap coordinates give geometric sums directly. The equalities then hold as rational functions by the identity theorem, and evaluation at a common regular point preserves them.
\end{proof}

The sum over all times is not interpreted as a convergent bilateral series at $\zeta$. To justify its value, attach independent formal character variables to the relative-time coordinates (equivalently to all nonroot edge times). Each choice of local ordering region, including the lower-dimensional boundary regions, gives a relatively open rational polyhedral cone in these coordinates. Its integer-point generating function with the local weight is a rational Laurent generating function: on each cone the weight is a character, and the positive-gap sum is a finite product of geometric factors. The product of the local cones is pointed because each local cone is pointed. Subdivide this cone according to the full ordering of the edge times, including equality strata. This subdivision preserves the rational generating function by Lemma~\ref{lem-cones}. The character variables allow the convergence domain to be chosen independently of the root-of-unity weights.

More explicitly, take the lattice $\Z^{d-1}$ with the nonroot edge times as coordinates. The map to child-minus-parent times is unimodular: its inverse sums differences along the unique path of parent edges, and both maps have integer coefficients. A weak total ordering of all edge times is a unique ordered partition into equality blocks; restricting to the root time zero leaves one independent strictly positive integral gap for each pair of adjacent blocks. The sign of each difference between two incident edges is fixed by this ordering, so each full ordering cone lies in exactly one product of local ordering regions (including their equality strata). Conversely these full cones partition each such product, without overlaps or omitted boundaries. On a local product cone the weight is a single character, whose restriction to each full cone is the same global weight. Thus the valuation identity is applied separately on each pointed product cone; no common domain of absolute convergence across distinct product cones is needed.

The full ordering cones are exactly those used in \eqref{eq-blocks}. The transformation from relative times to edge times is an integral bijection; in a full ordering cone the independent positive gaps are its integer coordinates. At the specialization of the auxiliary character variables to one, their geometric denominators are $1-\zeta^{k(F_r)}$ in the global cones and $1-Q^m$, $1-Q^n$, $1-Q^{m+n-1}$ (or their inverse-power versions) in the local cones. All are nonzero: $1\leq k(F_r)\leq j-1$ and $1\leq m,n,m+n-1\leq j-1$. Hence the rational identities can be evaluated termwise at this specialization. Summing the subdivisions gives \eqref{eq-blocks}, while the independent local sums give the product of \eqref{eq-localcalc}. This proves Theorem \ref{thm-tree} without using an ill-defined bilateral sum.

\section{Connected normal form and square-free matching}\label{app-normal}
\subsection{Construction to arbitrary finite order}
\begin{lemma}[Connected transpose-preserving normal form]\label{lem-normal}
To any finite transverse degree there exists a holomorphic formal orthogonal matrix $O=\Id+O(t_e)$ for which
\begin{equation}\label{eq-nf}
 \mathsf T'=O^{\mathsf T}\mathsf T O=D^{1/2}e^{\mathcal A}D^{1/2},\qquad
 \mathcal A^{\mathsf T}=\mathcal A,\qquad
 \mathcal A_k=0\ \ (j\nmid k)
\end{equation}
through that degree. Every coefficient of $\mathcal A$ is connected and has vertex-parity covariance: conjugation by $(-1)^{\widehat n_v}$ is equivalent to changing the signs of all edge variables incident to $v$. The first-order number-preserving part is the hard-core nearest-neighbor hopping operator in each fixed particle-number sector; it has zero vacuum matrix element.
\end{lemma}
\begin{proof}
Fix a finite layer and write $\mathcal O_\zeta$ for the local ring of holomorphic germs in $s$ at the chosen primitive root (holomorphic functions identified when they agree on some neighborhood of $\zeta$), with a local square-root branch of $z$. Let $\mathfrak m=(t_e:e\in E(\Omega))$. At target degree $M$ work in the truncated ring
\begin{equation}\label{eq-ring}
 \mathcal R_M=\mathcal O_\zeta[t_e:e\in E(\Omega)]/\mathfrak m^{M+1}.
\end{equation}
All positive-degree matrix series are nilpotent in this ring; their exponential, logarithm and Baker--Campbell--Hausdorff (BCH) operations are finite. The generator at each degree is determined by the preceding degrees and holomorphic unit denominators. Thus the construction is compatible with the natural projections $\mathcal R_{M+1}\to\mathcal R_M$; no analytic convergence of the full normal-form transformation is required. An operator coefficient of a transverse monomial is called connected if it acts only on the vertices of the support graph of that monomial, and it vanishes if that support graph is disconnected. The series $V$ of \eqref{eq-transfer} is connected in this sense. The scalar factors in \eqref{eq-Tnorm} can also be included as connected one-edge terms in $V$ if desired.

If $A,B$ are connected series with zero constant term, their commutator has the same property. Terms with disjoint vertex supports commute, and otherwise the union of two connected intersecting supports is connected. Conjugation by $D^{\alpha}$ preserves local support since $D$ is a product of one-site factors. Projection to particle-number grade $k$ also preserves local support: for an operator supported on $S$, only the restriction of $\Nop$ to $S$ is relevant. All these operations preserve covariance under the diagonal bit-parity operators at the vertices, together with the corresponding sign changes of incident transverse variables. Explicitly, conjugation by $(-1)^{n_v}$ is equivalent to $t_e\mapsto-t_e$ for $e$ incident to $v$. Therefore an operator coefficient at monomial $\prod_e t_e^{m_e}$ can change an occupation set only by the symmetric difference of the odd-multiplicity edge endpoints.

Inductively suppose the transfer matrix has the form $D^{1/2}e^{\mathcal A}D^{1/2}$, with $\mathcal A^{\mathsf T}=\mathcal A$ and connected coefficients, and that all nonresonant grades have been removed through degree $r-1$. Write $\mathcal A^{(r)}_k$ for the homogeneous degree-$r$ part of grade $k$. Choose
\begin{equation}
 S^{(r)}=-\sum_{j\nmid k}\frac{\mathcal A^{(r)}_k}{z^k-z^{-k}}.
\end{equation}
Since transposition sends grade $k$ to grade $-k$, $S^{(r)}$ is skew-symmetric. In
\begin{equation}
 e^{-S^{(r)}}D^{1/2}e^{\mathcal A}D^{1/2}e^{S^{(r)}}
 =D^{1/2}\left(e^{-D^{-1/2}S^{(r)}D^{1/2}}e^{\mathcal A}e^{D^{1/2}S^{(r)}D^{-1/2}}\right)D^{1/2},
\end{equation}
the new degree-$r$ logarithm of the middle product is $\mathcal A^{(r)}_k+(z^k-z^{-k})S^{(r)}_k$ in grade $k$. All nonresonant parts at that degree vanish. Its logarithm is symmetric because the middle product is symmetric and has identity constant term. The Baker--Campbell--Hausdorff expansion consists of the individual logarithms and their nested commutators, so the support and parity covariance properties persist. Denominators of the generators are holomorphic units near $s=\zeta$; each step involves finitely many grades. The last assertion follows because $\operatorname{arctanh}t_e=t_e+O(t_e^3)$ and a two-bit flip preserves number exactly when one endpoint is occupied and the other is not. At a fixed finite target degree, one may discard all higher terms of $\mathcal A$ and exponentiate the resulting polynomial series; the resulting matrix is exactly block diagonal in number modulo $2j$ and agrees with the original conjugate matrix through the target degree. This replacement does not change the desired rational perturbation coefficients, which are determined by matching degrees at generic $s$.
\end{proof}

\subsection{Minimal source degree and tree matching}\label{app-match}
Consider a matrix element from the vacuum to a set $S$ of $2mj$ particles, $m\geq1$. A connected coefficient of degree $r$ in $\mathcal A$ can involve at most $r+1$ sites, so its creation degree is at least $2mj-1$. In particular the source from the vacuum to the first resonant sector of $2j$ particles has degree at least $d=2j-1$. The lower bound $d$ for the first nonzero vacuum source also holds in $e^{\mathcal A}$: its terms are products of resonant factors, and the first factor that takes the vacuum to any nonzero number sector already has degree at least $d$. At order $d$ only the single creation term can contribute; all other factors have positive degree. For a target of $2j$ particles, equality of the connected-support lower bound forces all vertices of the support to be occupied, the edges to be distinct, and the support to be a tree. The parity covariance forces all degrees odd, hence degrees one or three on the square lattice.

\begin{lemma}[Square-free resonance matching]\label{lem-matching}
In the normal form of Lemma~\ref{lem-normal}, the coefficient $B_{\mathcal T}$ of the full square-free monomial of an embedded odd-degree tree $\mathcal T$ on $2j$ vertices, in the vacuum-to-$V(\mathcal T)$ matrix element of $\mathsf T'$, is $z^{-j}g_j(\mathcal T;\zeta)$ at $s=\zeta$. Here the root and the local square-root branch are chosen as above.
\end{lemma}
\begin{proof}
We match the actual coefficient to \eqref{eq-rec}, rather than identifying it by a choice of effective-Hamiltonian convention. Fix one such tree with edge set $E$, switch off all other variables, and work in the square-free algebra $t_e^2=0$ for $e\in E$. At $s=\zeta$, the only square-free monomials that can take the vacuum to the resonant sectors are the empty monomial and the full monomial $t_E=\prod_{e\in E}t_e$. The reason is that $F\mapsto\partial F$ is injective and $|\partial F|$ is strictly between zero and $2j$ for nonempty proper subsets. The vacuum diagonal is identically one in this algebra, by the same injectivity. Denote the full tree coefficient of the resonant matrix element of $\mathsf T'$ by $B_{\mathcal T}$. Thus
\begin{equation}\label{eq-nftree}
 \mathsf T'\vac=\vac+B_{\mathcal T} t_E\ket{V(\mathcal T)}.
\end{equation}
In this square-free algebra $e^V=\prod_{e\in E}(\Id+t_eX_e)$, so the analytic scalar normalization is irrelevant. Pass to the nonsymmetric matrix $\mathsf T_{\rm ns}=D\prod_{e\in E}(\Id+t_eX_e)$ and put $v=D^{1/2}O\vac$. Since $\mathsf T_{\rm ns}=D^{1/2}\mathsf T D^{-1/2}$, Eq.~\eqref{eq-nftree} gives
\begin{equation}\label{eq-veq}
 \mathsf T_{\rm ns}v=v+z^j B_{\mathcal T} t_E\ket{V(\mathcal T)}.
\end{equation}
Here all positive-degree corrections to $O\ket{V(\mathcal T)}$ vanish upon multiplication by $t_E$. Parity covariance and the square-free condition imply that $v$ has the expansion
\begin{equation}
 v=\sum_{F\subseteq E}v_F t_F\ket{\partial F},\qquad v_\varnothing=1.
\end{equation}
Comparing the proper-subset coefficients in \eqref{eq-veq} gives $v_F=q(F)$ by \eqref{eq-rec}. At the full edge set, the diagonal factor is $s^j=1$, so the unknown $v_E$ cancels and
\begin{equation}
 z^jB_{\mathcal T}=\sum_{F\subsetneq E}q(F)=g_j(\mathcal T;\zeta).
\end{equation}
This proves the claimed matching, including the common sign from the transfer-matrix convention. All transformations used here are the same edge-local transformations as in Lemma \ref{lem-normal}; the restriction to tree variables and the square-free algebra commutes with their finite-order construction. Thus the result is the actual monomial coefficient of the full normal form, not a separately chosen tree gauge.
\end{proof}

\section{Higher pole orders and the hard-core resolvent}\label{app-higher}
For integer coefficient indices $m,n\geq0$, let $p_j(m,n)$ be the pole order of $r_{mn}(s)$ at a primitive $j$th root, with value zero for a regular coefficient. This is independent of the primitive root by rationality over $\Q$.
\begin{theorem}[Highest pole orders]\label{thm-orders}
For $j\geq2$ and all $m,n\geq0$,
\begin{equation}\label{eq-bound}
 p_j(m,n)\leq \max\{0,2m+2n-4j+3\}.
\end{equation}
For $m+n\geq d=2j-1$, equality holds if and only if
\begin{equation}\label{eq-budget}
 \min(m,n)\geq c_j:=\left\lceil\frac{j-1}{2}\right\rceil.
\end{equation}
In this range, the budget condition is equivalent to the existence of an embedded odd-degree tree with $2j$ vertices having at most $m$ horizontal and $n$ vertical edges, respectively. If the budget condition fails, then $p_j(m,n)\leq 2m+2n-4j+2$. In particular, on the first-resonance line $m+n=d$, the pole order is one precisely when \eqref{eq-budget} holds and is zero otherwise.
\end{theorem}

We first state a stronger coefficientwise identity, proved in Appendix~\ref{app-schur}, and then deduce Theorem~\ref{thm-orders} using the embedding construction in Appendix~\ref{app-embedding}. Neither higher-pole result is needed for the first-pole proof in Section~\ref{sec-equal}.

\begin{proposition}[Leading-pole generating function]\label{prop-res}
Let $j\geq2$, let $\zeta$ be a primitive $j$th root, and set $\delta=1-s^j$ and $d=2j-1$. For $m+n\geq d$,
\begin{equation}\label{eq-limit}
 \lim_{s\to\zeta}\delta^{2m+2n-2d+1}r_{mn}(s)
 =\coef{X^{2m}Y^{2n}}\Hgen_j(X,Y),
\end{equation}
where
\begin{equation}\label{eq-resolvent}
 \Hgen_j(X,Y)=\lim_{\Omega\nearrow\Z^2}\frac1{|\Omega|}\,
 \Bsrc_j^{\mathsf T}(\Id-\mathsf A_j)^{-1}\Bsrc_j.
\end{equation}
The limit and inverse in \eqref{eq-resolvent} are coefficientwise formal power series. For $m+n<d$, $r_{mn}$ is regular at $\zeta$.
\end{proposition}
\begin{proof}[Proof idea]
Use $x=\delta X$, $y=\delta Y$ in the connected resonant normal form. The vacuum-to-$2j$ source is $\delta^d\Bsrc_j+O(\delta^{d+1})$. The gap in the $2j$ sector and its hopping both have order $\delta$, so the inverse gap is $\delta^{-1}(\Id-\mathsf A_j)^{-1}$ at leading order. The Schur complement therefore contributes $\delta^{2d-1}\Bsrc_j^{\mathsf T}(\Id-\mathsf A_j)^{-1}\Bsrc_j$ to the vacuum eigenvalue. Higher resonant sectors need higher-order connected sources; their first-order blocks do not mix particle numbers. The vacuum diagonal part is holomorphic and cannot contribute to the highest poles at the orders under consideration. Appendix \ref{app-schur} proves this as a statement about formal series with coefficients holomorphic in $\delta$, without assuming an analytic scaling limit at fixed $X,Y$ in infinite volume. The coefficientwise limit in \eqref{eq-resolvent} follows from finite support of each source and each finite sequence of hops.
\end{proof}

\begin{proof}[Proof of Theorem~\ref{thm-orders}]
At the principal root, all monomials in all $\Bsrc_{j,S}$ have the same real sign by Corollary \ref{cor:sign}. The expansion
\begin{equation}\label{eq-walk}
 \Hgen_j=\lim_{\Omega\nearrow\Z^2}\frac1{|\Omega|}
 \sum_{\ell\geq0}\Bsrc_j^{\mathsf T}\mathsf A_j^\ell\Bsrc_j
\end{equation}
has nonnegative coefficients. If a tree satisfies the two direction budgets, creating its occupied set, performing the required pairs of backtracking hops in the two directions, and annihilating the same set gives a strictly positive contribution at $X^{2m}Y^{2n}$. The positivity persists per unit area by translating that finite configuration into the bulk. Conversely, every trivalent vertex requires at least one edge in each transverse direction, and a given edge can serve at most two trivalent vertices. Hence each tree has $n_x,n_y\geq c_j$. Every entry of $\Bsrc_j$ is divisible by $X^{c_j}Y^{c_j}$, and every coefficient of the resolvent contribution below either doubled budget is zero. Appendix~\ref{app-embedding} constructs trees for all admissible budgets and gives the backtracking argument in detail. Proposition~\ref{prop-res} supplies the bound and identifies its leading coefficient; the preceding positivity and divisibility arguments give the equality criterion and the improved bound when a budget fails. Galois conjugation transfers the pole orders from $\zeta_j$ to every primitive $j$th root.
\end{proof}

\section{Schur complement and coefficientwise highest poles}\label{app-schur}
\begin{proof}[Detailed proof of Proposition~\ref{prop-res}]
For a finite layer, fix a maximal transverse degree $M$ and use the polynomial normal form of Lemma \ref{lem-normal} through that degree. Choose the local coordinate $\delta=1-s^j$, with $s$ the branch through $\zeta$. Restrict to the resonant block containing the vacuum, namely the direct sum of number sectors $0,2j,4j,\ldots$. At transverse variables $x=\delta X$, $y=\delta Y$, this block of $\mathsf T'$ equals the identity at $\delta=0$. Consequently
\begin{equation}\label{eq-scaled}
 \mathsf C(\delta,X,Y)=\frac{\mathsf T'_{\rm res}-\Id}{\delta}
\end{equation}
is holomorphic in $\delta$ (for the finite matrix and its formal transverse coefficients). At $X=Y=0$ the diagonal of $\mathsf C$ in number sector $2mj$ is
\begin{equation}
 \frac{(1-\delta)^m-1}{\delta}=-m+O(\delta),\qquad m\geq0.
\end{equation}
The vacuum eigenvalue branch $\ell$ of $\mathsf C$ is therefore defined by a formal expansion in $X,Y$ whose coefficients are holomorphic in $\delta$: the vacuum gap at $X=Y=\delta=0$ is nonzero. This follows by the same coefficient recursion as in Appendix \ref{app-transfer}, now with holomorphic unit denominators. In particular $\lambda=1+\delta\ell$ is the original vacuum eigenvalue after the substitution, coefficientwise, by uniqueness at generic $s$.

At $\delta=0$, the first-order hopping preserves particle number and has zero vacuum entry. The scaled block is block diagonal by particle number with diagonal blocks $-m\Id+\mathsf A_{mj}$, where $\mathsf A_{mj}$ is the linear hard-core hopping operator in the $2mj$ sector. Hence $\ell(0,X,Y)=0$ as a formal series. Write $\mathsf C$ in vacuum/complement blocks as
\begin{equation}
 \mathsf C=\begin{pmatrix}c_{00}&B_+^{\mathsf T}\\B_+&C_+\end{pmatrix}.
\end{equation}
By the minimal-source lemma, $B_+=\delta^{d-1}\widehat B+O(\delta^d)$, where $\widehat B$ equals $\Bsrc_j(X,Y)$ in the $2j$ sector and is zero in the other sectors. The zero statement at degree $d$ for higher sectors follows because a connected single source to $2mj>2j$ particles requires more than $d$ edges; products of two positive-degree terms cannot contribute at this first creation order. All error statements here and below refer to formal $X,Y$ coefficients with holomorphic $\delta$ coefficients. The Schur complement equation is
\begin{equation}\label{eq-Schur}
 \ell=c_{00}+B_+^{\mathsf T}(\ell\Id-C_+)^{-1}B_+.
\end{equation}
The inverse exists in the formal ring because its constant term at $X=Y=\delta=0$ is diagonal with strictly positive entries $m$. At $\delta=0$, its restriction to the $2j$ sector is $(\Id-\mathsf A_j)^{-1}$. Therefore
\begin{align}
 \lambda-\lambda_{00}&=\delta(\ell-c_{00})\notag\\
 &=\delta^{2d-1}\Bsrc_j^{\mathsf T}(\Id-\mathsf A_j)^{-1}\Bsrc_j+O(\delta^{2d}),\label{eq-lambdaPole}
\end{align}
where $\lambda_{00}=1+\delta c_{00}$ is the vacuum diagonal entry of $\mathsf T'$ at the scaled couplings. In particular $\lambda_{00}(0,X,Y)=1$. Taking logarithms does not alter the displayed leading term in the difference $\log\lambda-\log\lambda_{00}$.

Before scaling, $\lambda_{00}$ is holomorphic near $s=\zeta$ in each transverse coefficient. Its logarithm has constant transverse coefficient zero, and its homogeneous degree-$M'$ term acquires a factor $\delta^{M'}$ on scaling. For degrees $M'\geq2d$, the leading term of \eqref{eq-lambdaPole} therefore gives the highest possible pole order $M'-(2d-1)$. At degrees $M'<2d$, the original (unscaled) off-diagonal vacuum blocks have transverse order at least $d$; the Schur-complement correction to the vacuum eigenvalue begins at order $2d$ because the inverse of the complementary block has transverse order zero at generic $s$. Hence all smaller-degree terms come only from the holomorphic vacuum diagonal. This proves the regularity part as well as the bound \eqref{eq-bound} in finite volume.

Take $M'=2m+2n$ and divide \eqref{eq-lambdaPole} by $|\Omega|$. At fixed degree, Appendix \ref{app-transfer} ensures that the thermodynamic rational coefficient is a finite sum of connected translation classes. On the right-hand side, a creation tree and a fixed number of hard-core hops occupy a bounded neighborhood of the initial tree; a final annihilation tree must have the resulting occupied set as its vertex set. Thus the coefficient of the resolvent in \eqref{eq-resolvent} is also a finite sum per site with only vanishing boundary errors. Taking these stabilized rational coefficients and their Laurent coefficients gives \eqref{eq-limit}. This argument never assumes uniform convergence in $s$ near the pole for the infinite-volume partially resummed series.
\end{proof}

\section{Direction budgets and nonzero coefficients}\label{app-embedding}
Let $r=j-1$ be the number of trivalent core vertices. Necessity of the directional budget follows from counting incidences: every trivalent vertex in the square lattice has at least one horizontal and one vertical edge, and each horizontal or vertical edge contributes to at most two such incidences. Thus $2n_x\geq r$ and $2n_y\geq r$. To prove sufficiency of the required embedding criterion, it suffices to construct chain-core trees with transverse edge numbers $(n_x,n_y)$ satisfying
\begin{equation}\label{eq-interval}
 n_x+n_y=2r+1,
 \qquad \left\lceil\frac r2\right\rceil\leq n_y\leq 2r+1-\left\lceil\frac r2\right\rceil.
\end{equation}
For $r=1$, the three-edge stars give $n_y=1,2$. For general $r$, first choose $0\leq k_0\leq\lfloor r/2\rfloor$. Place $k_0$ consecutive vertical dimers in a northeast staircase: their lower and upper vertices are $(i,i)$ and $(i,i+1)$, $i=0,\ldots,k_0-1$. Join the upper vertex of each dimer to the lower vertex of the next by a horizontal edge. Append the remaining $r-2k_0$ core vertices on the horizontal ray with coordinates $(k_0+k,k_0)$, $k=0,\ldots,r-2k_0-1$, joined to the last upper vertex if $k_0>0$. If $k_0=0$, use the ordinary horizontal chain $(k,0)$ instead.

For each dimer lower vertex attach a horizontal leaf on its right and for each upper vertex attach a horizontal leaf on its left; these are the unused horizontal directions at the internal bends. At the first lower vertex also attach a left leaf to make its degree three, and if the core ends at an upper vertex attach an extra right leaf there. At each appended single vertex attach a leaf immediately above it; the final endpoint also gets a horizontal leaf on the right. For the all-horizontal chain attach the same upper leaves and the necessary exterior horizontal endpoint leaves. The one-vertex case is interpreted as a star. These coordinates are distinct: the right leaves of lower dimer vertices lie one row below the subsequent core bend, the left leaves of upper vertices lie to the left of its core column, and the appended upper leaves lie beyond the staircase row. Thus the chosen edges form an embedded tree (the lattice may contain unused edges between its vertices; an induced tree is not required).

The only vertical edges are the $k_0$ core dimer edges and the $r-2k_0$ leaves at single vertices. This gives $n_y=r-k_0$ and $n_x=2r+1-n_y$, realizing the interval from $\lceil r/2\rceil$ to $r$. Exchanging the two axes gives the remaining integers from $r+1$ to $2r+1-\lceil r/2\rceil$. This proves \eqref{eq-interval}.

Suppose $m+n\geq2r+1$ and $\min(m,n)\geq\lceil r/2\rceil$. The interval \eqref{eq-interval} contains an integer $n_y$ with $n_y\leq n$ and $n_x=2r+1-n_y\leq m$; this is the intersection of $[\lceil r/2\rceil,2r+1-\lceil r/2\rceil]$ with $[2r+1-m,n]$. Choose the corresponding embedded tree. For either transverse direction, a finite occupied set has an extremal site adjacent to an unoccupied site in that direction. One may move the particle there and immediately back. These two hops return the occupied set to itself and supply a factor $X^2$ or $Y^2$. Repeat $m-n_x$ and $n-n_y$ times, respectively. The contribution using the same initial and final tree has the required monomial $X^{2m}Y^{2n}$ and is nonzero, with a positive sign at the principal primitive root. Since every term of \eqref{eq-walk} at that root is nonnegative, the coefficient is strictly positive in the thermodynamic limit. When $m+n\geq d$ but either budget fails, $X^{2c_j}Y^{2c_j}$ divides the entire formal resolvent expression by the incidence count, so its coefficient at $X^{2m}Y^{2n}$ vanishes. Proposition~\ref{prop-res} then lowers the pole-order bound by at least one. When $m+n<d$, regularity has already been proved. This completes the justification of Theorem~\ref{thm-orders}.

\section{Computational consistency checks and reproducibility}\label{app-check}
Although the proof is analytic, the product formula can be checked in finite fields. For a prime $p\equiv1\pmod j$, let $\mathbb F_p$ be the field of integers modulo $p$ and choose an element of exact order $j$. All denominators of \eqref{eq-rec} and \eqref{eq-symvertex} are then invertible, so the subset recursion and the product can be evaluated without floating-point cancellation. We enumerate the nonisomorphic core trees of maximum degree three and add leaves to complete each core vertex to degree three. Table~\ref{tab-treecheck} reports the comparison of the subset recursion with the symmetric vertex product \eqref{eq-symvertex}.
\begin{table}[htbp]
\centering
\caption{Exact finite-field verification of
the resonant tree factorization formula.
For each resonance order $j$, the subset
recursion and the factorized expression are
evaluated in $\mathbb F_p$.}
\label{tab-treecheck}
\begin{tabular}{rrrrl}
\toprule
$j$ & Prime $p$ & Root in $\mathbb F_p$
& Core topologies & Result\\
\midrule
2 & 2003 & 2002 & 1 & agree\\
3 & 3001 & 934  & 1 & agree\\
4 & 4001 & 3102 & 1 & agree\\
5 & 5011 & 273  & 2 & agree\\
6 & 6007 & 5930 & 2 & agree\\
\bottomrule
\end{tabular}
\end{table}
For the same five finite fields and all listed core topologies, the updated script also computes the unrooted internal-edge formula \eqref{eq-edgeprod} directly, and checks the shifted Cayley composition \eqref{eq-shiftedmob} and determinant formula \eqref{eq-detvertex} for every admissible pair $m,n\geq1$, $m+n\leq j$. These are independent algebraic implementations of the identities; all checks agree modulo the indicated primes. These checks are supplementary and are not used to establish the all-order statement.

As a separate normalization check, a direct symbolic eigenvector recursion for a single three-edge star with uniform edge variable $t$ gives
\begin{equation}
 [t^6]\log\lambda_{\rm star}(t,s)=
 -\frac{s^2(s^4-4s^3-9s^2-22s-14)}{(s-1)^5(s+1)}.
\end{equation}
Multiplication by $1-s^2$ and evaluation at $s=-1$ gives $1/4$, in agreement with the squared tree amplitude and hence with the density formula \eqref{eq-star}. The direct computation uses the normalized product transfer matrix of Appendix \ref{app-transfer}.

\subsection{Independent finite-tree transfer-matrix checks}
A further check works directly with the nonsymmetric transfer matrix $D\prod_e(\Id+\delta\xi X_e)$ on a single finite tree, with $s=\zeta(1-\delta)^{1/j}$, over $\mathbb F_p[[\delta]]$. Here $\xi$ is a scalar formal variable. The occupation basis is indexed by tree-edge subsets via the injective boundary map. For a subset $F$, put $D_F=s^{|\partial F|/2}$, and expand the vacuum-normalized eigenvector and eigenvalue as $v_F=\sum_n v_{n,F}\xi^n$ and $\lambda=\sum_n\lambda_n\xi^n$. Set $v_{0,F}=\mathbf1_{F=\varnothing}$, $v_{n,\varnothing}=0$ for $n>0$, and $\lambda_0=1$. Define
\begin{equation}
 U_{n,F}=\sum_{\substack{H\subseteq E(\mathcal T)\\1\leq |H|\leq n}}\delta^{|H|}v_{n-|H|,F\triangle H}.
\end{equation}
The recursion, without any normal-form transformation, is
\begin{equation}
 \lambda_n=U_{n,\varnothing},\qquad
 (1-D_F)v_{n,F}=D_F U_{n,F}-\sum_{k=1}^{n-1}\lambda_k v_{n-k,F}\quad(F\ne\varnothing).
\end{equation}
The logarithm is obtained by formal differentiation of $\lambda=\exp(\log\lambda)$. With $d=2j-1$, the computation retains $4d+4$ powers of $\delta$ and $2d$ powers of $\xi$. For these trees the only nonempty resonant subset is the full tree, for which $1-D_F=\delta$. At most one digit of $\delta$-adic precision is lost per transverse recursion step, leaving more than the $2d$ digits needed for the check. The primes are larger than all integer denominators from these truncated binomial and logarithmic series.

Table~\ref{tab-finitetree} compares the resulting first normalized pole amplitudes with the squared amplitudes predicted by Theorem~\ref{thm-tree} (using the matching of Lemma~\ref{lem-matching}).
\begin{table}[htbp]
\centering
\caption{Independent finite-tree transfer-matrix
verification of the first normalized pole
amplitudes. All entries are evaluated exactly
modulo the indicated prime $p$.}
\label{tab-finitetree}
\begin{tabular}{rrrr}
\toprule
$j$ & $p$
& $[\delta^{2d-1}\xi^{2d}]\log\lambda$
& $g_j(\mathcal T;\zeta)^2\pmod p$\\
\midrule
2 & 2003 & 501  & 501\\
3 & 3001 & 2853 & 2853\\
4 & 4001 & 3001 & 3001\\
\bottomrule
\end{tabular}
\end{table}
For these cases the script also checks that lower transverse orders have no negative Laurent powers before scaling and that the order $2d$ pole is at most simple. These are independent finite-tree transfer-matrix tests, not independent computations of the full thermodynamic lattice coefficients.

\subsection{Independent checks on a transverse graph with cycles}
We also compute directly on the open $3\times3$ square layer (nine sites, twelve edges, four elementary plaquettes). The even occupation sector has dimension $256$. For a general edge subset $H$ of this graph, the boundary map need not be injective. Index the eigenvector by even occupation sets $S$ rather than by edge subsets, and use
\begin{equation}
 U_{n,S}=\sum_{h=1}^{\min(n,12)}\delta^h
 \sum_{\substack{H\subseteq E(\Omega)\\|H|=h}}
 v_{n-h,S\triangle\partial H}.
\end{equation}
The eigenvalue and eigenvector recurrences are otherwise identical to the preceding subsection with $D_S=s^{|S|/2}$. All subsets are retained, including cycles, and equal boundaries are aggregated with their full multiplicities. The computations use $P=2M+6$ retained $\delta$-adic digits through transverse degree $M$. Each nonvacuum denominator has valuation zero or one (the primes exceed all particle counts and $M$); after factoring out a possible $\delta$ there is at most one digit of precision loss per recursion degree. Thus coefficients at powers below $P-M$ are exact over $\mathbb F_p$.

Let $B=\Bsrc_j(1,1)$ be the finite-layer creation vector, and let $A=\mathsf A_j(1,1)$. For an even total transverse degree $M'\geq 2d$, Proposition~\ref{prop-res} predicts, without dividing by layer area,
\begin{equation}
 [\delta^{2d-1}\xi^{M'}]\log\lambda_\Omega(\delta\xi,\delta\xi,s)
 =B^{\mathsf T}A^{M'-2d}B.
\end{equation}
Here $s=\zeta(1-\delta)^{1/j}$. In the prediction the common sign $z^{-j}$ may be omitted since it squares to one. The direct recursion does not use either the tree factorization or the normal form. The prediction is computed separately by enumerating connected odd-degree tree edge sets on the layer and by applying the nearest-neighbor hopping matrix repeatedly. Table~\ref{tab-gridcheck} reports exact modular agreement, including several orders beyond the first pole. The program also checks vanishing of all coefficients at $\delta$ powers below the indicated bound, and regularity before degree $2d$.
\begin{table}[htbp]
\centering
\caption{Direct full-layer transfer recursion versus the hard-core resolvent on the open $3\times3$ grid. All amplitudes are unnormalized by area and reduced modulo $p$.}\label{tab-gridcheck}
\begin{tabular}{rrrrrr}\toprule
$j$ & $p$ & $M'$ & $2d-1$ & Direct & Predicted\\\midrule
2 & 2003 & 6 & 5 & 2 & 2\\
2 & 2003 & 8 & 5 & 21 & 21\\
2 & 2003 & 10 & 5 & 616 & 616\\
2 & 2003 & 12 & 5 & 461 & 461\\
3 & 3001 & 10 & 9 & 1225 & 1225\\
3 & 3001 & 12 & 9 & 2022 & 2022\\
4 & 4001 & 14 & 13 & 1 & 1\\\bottomrule
\end{tabular}
\end{table}
For clarity, the last row of Table~\ref{tab-gridcheck}
($j=4$, $p=4001$) can also be checked by a small
characteristic-zero enumeration. Here $d=7$ and the
coefficient is $[\delta^{13}\xi^{14}]\log\lambda_\Omega$.
There are exactly four connected seven-edge odd-degree
trees on the open $3\times3$ layer, one for each choice
of the omitted noncorner boundary site. Their eight-site
occupation sets are distinct. Each has a three-vertex
path as its trivalent core and amplitude $g_4=-1/2$
at the principal root by \eqref{eq-chain}.
Consequently the unnormalized quadratic amplitude is
$4(1/4)=1$. In the field $\mathbb F_{4001}$ the primitive
fourth root used is $3102$, with $3102^2\equiv-1\pmod{4001}$.
A single-tree square is $1/4\equiv3001\pmod{4001}$,
as in Table~\ref{tab-finitetree}, whereas the four-tree
sum is $4\cdot3001\equiv1\pmod{4001}$.
This finite-layer sum is distinct from the infinite-layer
amplitude $R_4(1)=19$ in Table~\ref{tab-amplitudes}.
The companion script verifies the four-tree count and
the exact rational sum separately.

These are tests of the equal-transverse finite-layer coefficients, not an independent proof of the infinite-volume theorem or of all individual direction-resolved pole amplitudes. In particular, a modular check cannot exclude an arbitrary characteristic-zero discrepancy divisible by the chosen prime.

\subsection{Lattice enumeration}
To obtain Table~\ref{tab-amplitudes}, start with the two one-edge translation classes and recursively attach a nearest-neighbor edge ending at a previously absent vertex. Canonicalize by translation to the minimal horizontal and vertical coordinates and sort the resulting undirected edges. This generates all finite embedded trees up to the chosen edge count: removal of a leaf reduces any tree to one at the preceding step. Retain precisely the odd-degree trees, and for each vertex set collect its directional edge-count histogram. The quadratic expression \eqref{eq-Rj} at $a=1$, or its bivariate version \eqref{eq-firstpoly}, is then evaluated by exact rational arithmetic using \eqref{eq-chain} for $j\leq4$. The companion Python 3 script (NumPy is used only in the direct transfer-matrix check) implements the tree, full-layer and lattice-enumeration tests. No numerical test replaces any of the analytic proofs.

\clearpage
\section{Logical structure of the proofs}\label{app-roadmap}
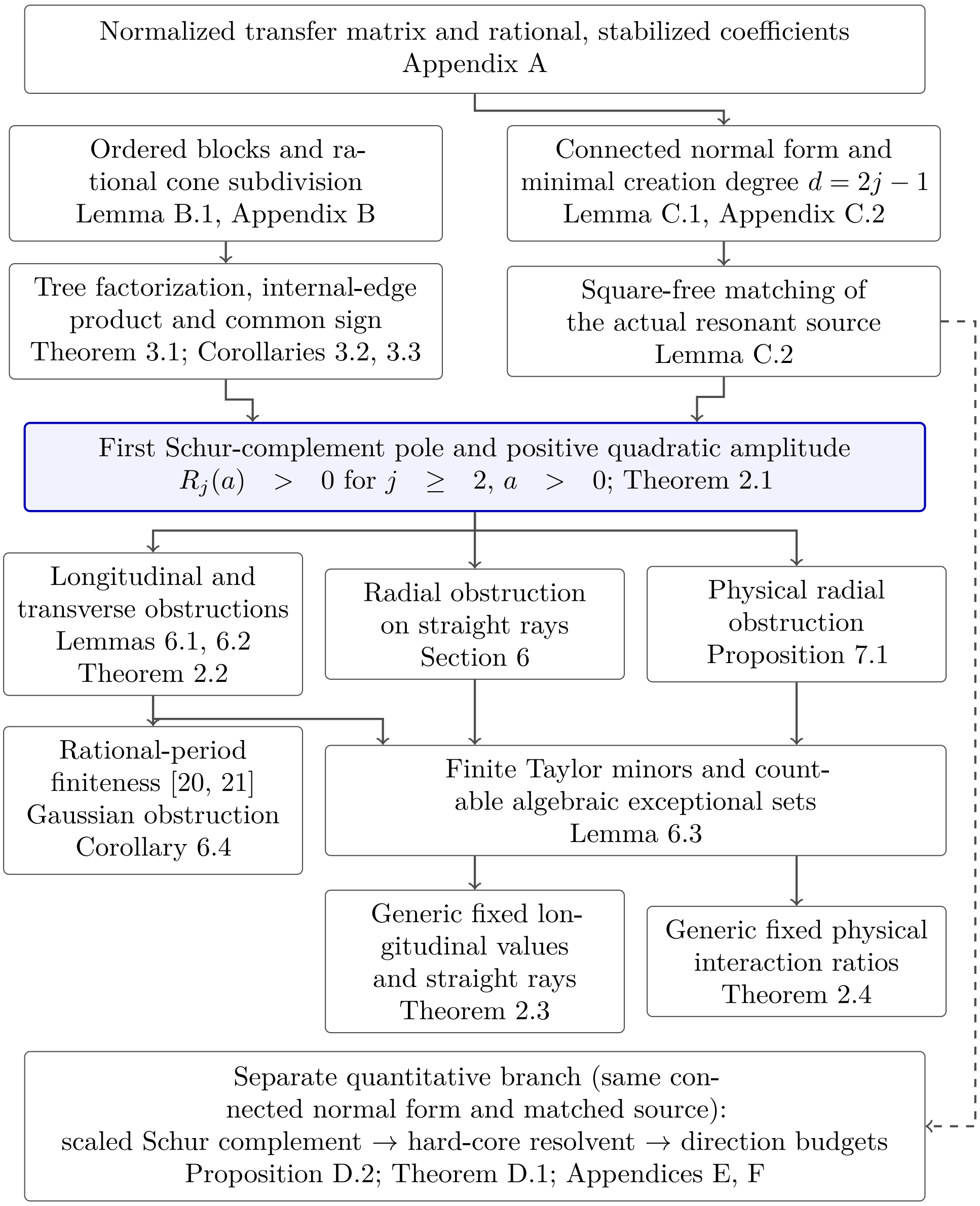
\begin{figure}[!htbp]
\centering
\begin{tikzpicture}[
 box/.style={draw=black!70,rounded corners=2pt,align=center,inner sep=5pt,font=\small,text width=5.3cm,minimum height=1.15cm},
 wide/.style={box,text width=11.4cm},
 narrow/.style={box,text width=3.55cm,font=\small},
 core/.style={wide,draw=blue!70!black,fill=blue!5,thick},
 arr/.style={->,thick,draw=black!70},
 minor/.style={arr,dashed}
 ]
\node[wide] (rational) at (0,0) {Normalized transfer matrix and rational, stabilized coefficients\\ Appendix~\ref{app-transfer}};
\node[box] (tree) at (-3.25,-1.75) {Ordered blocks and rational cone subdivision\\ Lemma~\ref{lem-cones}, Appendix~\ref{app-cones}};
\node[box] (nf) at (3.25,-1.75) {Connected normal form and minimal creation degree $d=2j-1$\\ Lemma~\ref{lem-normal}, Appendix~\ref{app-match}};
\node[box] (sign) at (-3.25,-3.55) {Tree factorization, internal-edge product and common sign\\ Theorem~\ref{thm-tree}; Corollaries~\ref{cor:edges}, \ref{cor:sign}};
\node[box] (match) at (3.25,-3.55) {Square-free matching of the actual resonant source\\ Lemma~\ref{lem-matching}};
\node[core] (first) at (0,-5.45) {First Schur-complement pole and positive quadratic amplitude\\ $R_j(a)>0$ for $j\geq2$, $a>0$; Theorem~\ref{thm-first}};
\node[narrow] (dir) at (-4.2,-7.5) {Longitudinal and transverse obstructions\\ Lemmas~\ref{lem-pole}, \ref{lem-transverse}\\ Theorem~\ref{thm-main}};
\node[narrow] (rays) at (0,-7.5) {Radial obstruction on straight rays\\ Section~\ref{sec-dfinite}};
\node[narrow] (phys) at (4.2,-7.5) {Physical radial obstruction\\ Proposition~\ref{prop-radial}};
\node[narrow] (gauss) at (-4.2,-9.8) {Rational-period finiteness~\cite{grothendieck1966,lairez2016}\\ Gaussian obstruction\\ Corollary~\ref{cor-gaussian}};
\node[box,text width=7.75cm] (minors) at (2.1,-9.8) {Finite Taylor minors and countable algebraic exceptional sets\\ Lemma~\ref{lem-generic}};
\node[narrow] (spec) at (0,-11.9) {Generic fixed longitudinal values and straight rays\\ Theorem~\ref{thm-special}};
\node[narrow] (physical) at (4.2,-11.9) {Generic fixed physical interaction ratios\\ Theorem~\ref{thm-physical}};
\node[wide] (higher) at (0,-14.05) {Separate quantitative branch (same connected normal form and matched source):\\ scaled Schur complement $\to$ hard-core resolvent $\to$ direction budgets\\ Proposition~\ref{prop-res}; Theorem~\ref{thm-orders}; Appendices~\ref{app-schur}, \ref{app-embedding}};
\draw[arr] (rational.south) -- ++(0,-.22) -| (nf.north);
\draw[arr] (tree) -- (sign);
\draw[arr] (nf) -- (match);
\draw[arr] (sign.south) -- ++(0,-.26) -| ([xshift=-2.9cm]first.north);
\draw[arr] (match.south) -- ++(0,-.26) -| ([xshift=2.9cm]first.north);
\draw[arr] (first.south) -- ++(0,-.24) -| (dir.north);
\draw[arr] (first.south) -- (rays.north);
\draw[arr] (first.south) -- ++(0,-.24) -| (phys.north);
\draw[arr] (dir.south) -- (gauss.north);
\draw[arr] (dir.south) -- ++(0,-.30) -| ([xshift=-3.3cm]minors.north);
\draw[arr] (rays.south) -- ([xshift=-2.1cm]minors.north);
\draw[arr] (phys.south) -- ([xshift=2.1cm]minors.north);
\draw[arr] ([xshift=-2.1cm]minors.south) -- (spec.north);
\draw[arr] ([xshift=2.1cm]minors.south) -- (physical.north);
\draw[minor] (match.east) -- ++(.45,0) |- (higher.east);
\end{tikzpicture}
\caption{Proof dependencies. Solid arrows indicate the first-pole, Gaussian obstruction and specialization arguments; the dashed arrow indicates the separate higher-pole analysis using the same normal form and matched source. The intrinsic theorem needs neither the Taylor-minor argument nor the higher-pole resolvent. The isotropic specialization is not proved.}\label{fig-proofmap}
\end{figure}

Figure~\ref{fig-proofmap} summarizes the analytic dependencies. The first-pole theorem requires both the local tree calculation and the connected normal form; the latter alone does not exclude cancellation. Corollary~\ref{cor-gaussian} follows from Theorem~\ref{thm-main} and the finiteness of rational periods, independently of the generic specialization arguments and the higher-pole analysis. The higher-pole branch is independent of the differential-equation and specialization arguments. The finite-field checks in Appendix~\ref{app-check} are not premises of any theorem.

\clearpage

\section*{Acknowledgments}
The author wishes to express his sincere gratitude
to Professor Robert L. Bryant for his kindness
in corresponding with him and mailing him a copy
of his seminal paper on Willmore surfaces about
twenty years ago.
The geometric perspective of that work has
continued to inspire the author over the years
and ultimately helped motivate the present study.

\section*{Statements and Declarations}
\paragraph{AI use.} GPT-6.2 was used to assist in the exploration and checking of mathematical arguments, computational consistency checks, and manuscript preparation and revision. GLM-5.4 was additionally used for cross-checking mathematical arguments.

\paragraph{Data and code availability.} No external data sets are required. The accompanying script \texttt{ising\_resonance\_checks\_v8.py} reproduces the computational checks reported in Appendix~\ref{app-check}.

\end{document}